\documentclass[aps,prb,twocolumn,superscriptaddress]{revtex4-2}
\usepackage{amsmath,txfonts,braket}
\usepackage{graphicx}

\usepackage{bm,wrapfig}
\usepackage{epsfig}
\usepackage{amsfonts}
\usepackage{amssymb}
\usepackage{color}

\newcommand{\bs}[1]{\boldsymbol{#1}}
\newcommand{\pa}{\partial}

\newcommand{\ve}{\varepsilon}

\newcommand{\dket}{\rangle \rangle }

\begin{document}
\title{ 
Shockwave-Enhanced Floquet Engineering of Relativistic Quasiparticles
}
\author{Takashi Oka}
\affiliation{Institute for Solid State Physics, University of Tokyo, Kashiwa 277-8581, Japan}
\affiliation{Trans-scale Quantum Science Institute, University of Tokyo, Bunkyo-ku, Tokyo 113-0033, Japan}

\date{\today}

\begin{abstract}

We investigate Floquet engineering of relativistic quasiparticles driven by
propagating waves, identifying distinct quantum states and phase transitions
in the time-like, light-like, and space-like regimes. Using Floquet--Bloch
theory, we show that the Lorentz redshift of the driving frequency in the rest
frame of the wave strongly amplifies Floquet band modulation as the wave speed
approaches the Fermi velocity, leading to a shockwave-like state synchronized
with the wave motion. For two-dimensional Dirac fermions in an in-plane
circular gauge wave, the boost also enhances the longitudinal field, so that the
Floquet Dirac mass grows with the square of the Lorentz factor, and the two
helicities invert the mass on opposite sides of the light line; on the honeycomb
lattice this inversion is valley selective and drives a Floquet Chern
transition. For three-dimensional Dirac fermions, we uncover a regime where
Floquet Weyl bands emerge and transition into type-II Weyl states as the wave
speed nears the Fermi velocity. These findings extend beyond electrons to
quasiparticles with relativistic dispersions, with slow surface waves such as
acoustic waves, plasmons, and spin waves as natural drives, opening new avenues
for dynamic band engineering in quantum materials.

\end{abstract}

\maketitle

\section{Introduction}
\label{sec:introduction}
\begin{figure}[tb]
\centering
\includegraphics[width=0.9\linewidth]{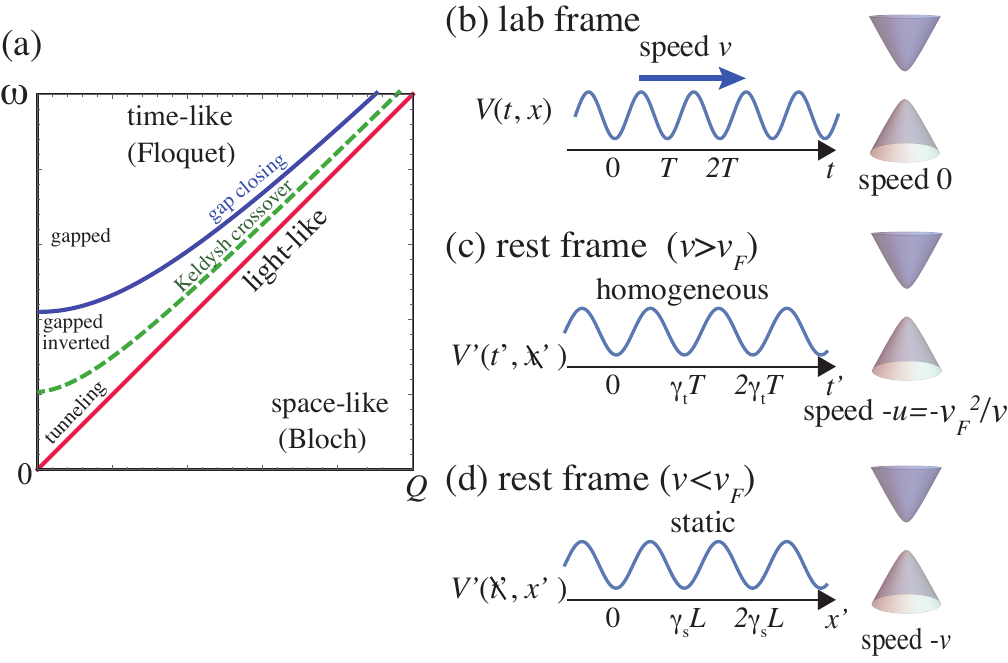}
\caption{
(a) Schematic phase diagram of a Dirac fermion driven by a propagating wave with wave number $Q$ and frequency $\omega$, drawn for the two-dimensional model of Sec.~\ref{sec:2D}.
The light-like line $\omega=v_FQ$ (red) separates time-like waves ($v>v_F$), whose rest-frame problem is a Floquet problem, from space-like waves ($v<v_F$), whose rest-frame problem is a Bloch problem.
On approaching the light line from the time-like side, the Lorentz redshift of the drive enhances the Floquet renormalization of the Dirac mass,
which changes sign along the gap-closing line (solid blue curve), separating the gapped phase from the gapped phase with inverted mass
[for the three-dimensional model of Sec.~\ref{sec:3D} the inverted phase is replaced by a Floquet Weyl semimetal].
Closer to the light line the Keldysh parameter of Eq.~(\ref{eq:Keldysh}) drops below unity (dashed green line, schematic; its position depends on the polarization and amplitude of the wave) and real carrier creation by tunneling sets in.
(b) In the laboratory frame the wave propagates with speed $v$ while the Dirac electrons are at rest.
(c) For a time-like wave ($v>v_F$), a Lorentz boost with velocity $u=v_F^2/v$ (Sec.~\ref{sec:Lorentz-framework}) leads to the rest frame in which the wave is spatially homogeneous and oscillates with the dilated period $\gamma_tT$, $\gamma_t=(1-v_F^2/v^2)^{-1/2}$, while the Dirac electrons move with velocity $-u=-v_F^2/v$: a conventional Floquet problem with a redshifted frequency.
(d) For a space-like wave ($v<v_F$), the boost with $u=v$ leads to the rest frame in which the wave is static with the dilated wavelength $\gamma_sL$, $\gamma_s=(1-v^2/v_F^2)^{-1/2}$, while the electrons move with velocity $-v$: a Bloch problem in a static superlattice. A light-like wave ($v=v_F$) has no rest frame.
}
\label{fig:concept}
\end{figure}

Floquet engineering~\cite{RevModPhys.89.011004,OkaKitamura2019,Rudner2020,RevModPhys.93.041002,RevModPhys.91.015006} 
enables dynamic control of quantum systems via periodic driving, facilitating the realization of topological phases~\cite{OkaAoki2009,Lindner2011,PhysRevB.84.235108}, Weyl semimetals~\cite{
Wang_2014,PhysRevB.93.155107,PhysRevLett.116.026805,Hubener2017,yoshikawa2022lightinducedchiralgaugefield,PhysRevB.96.041126,PhysRevResearch.6.L012027}, and non-equilibrium quantum states~\cite{Harper2020,Mori2023}.
While most studies focus on uniform driving fields, recent advances explore propagating waves as a tool for band engineering~\cite{Oliva-Leyva_2016,XuWu2018,Puebla_2020,Peng2022,yang2024quantizedacoustoelectricfloqueteffect,walicki2024floquetengineeringnearlyflat}.
Here, we show that Floquet states of quasiparticles with relativistic dispersion driven by propagating waves exhibit a regime of strong band modification
when the wave speed approaches the Fermi velocity, which we call the {\it shockwave limit} (see also the related shockwave physics discussed for graphene plasmons near the
quasi-relativistic resonance~\cite{Jablan2020,jablan2025quantumshockwavequasirelativisticresonance}).
This enhancement arises from the Lorentz redshift of the driving frequency,
$\omega\to\omega'=\omega/\gamma_t$, in the time-like rest frame.
Due to this redshift, the Floquet effective Hamiltonian, which governs the long-time behavior of the system~\cite{PhysRevB.84.235108,Mikami2016}, 
scales as
 \begin{align}
\sum_{m>0} \frac{1}{m\omega}[H_{-m},H_{m}]\to
\sum_{m>0}\frac{\gamma_t}{m\omega}[H_{-m},H_{m}],
 \label{eq:1st}
 \end{align}
where $H_m$ are the Fourier components of the Hamiltonian, $H(t)=\sum_m e^{-im\omega t}H_m$, so that $H_{m>0}$ ($H_{m<0}$) describes the absorption (emission) of $m$ quanta.
Since the redshift factor  $\gamma_t$ increases as the shockwave limit  is approached, 
various Floquet engineering effects such as the Floquet topological transition~\cite{OkaAoki2009} are expected to take place.
We note that this estimate applies when the drive amplitudes are invariant under the boost.
As we show below, a drive with a longitudinal gauge component must instead be transformed together with the scalar potential.
Figure~\ref{fig:concept}~(a) summarizes the resulting structure in the plane of the wave number $Q$ and the frequency $\omega$,
and Figs.~\ref{fig:concept}~(b--d) illustrate the rest-frame picture that underlies it.

The Lorentz redshift also brings Floquet band renormalization into competition with
real carrier creation.  The excitation mechanism is characterized by a
Keldysh parameter of the form
\begin{align}
\gamma_{\rm K}\sim
\frac{\omega'\Delta_{\rm eff}'}{e v_F E'},
\label{eq:Keldysh}
\end{align}
where $\Delta_{\rm eff}'$ and $E'$ are the effective gap and electric-field
amplitude in the {\it rest frame}.  The regimes $\gamma_{\rm K}\gg1$ and
$\gamma_{\rm K}\ll1$ correspond, respectively, to perturbative multiphoton
absorption and nonadiabatic tunneling, with a
crossover near $\gamma_{\rm K}\sim1$~\cite{Keldysh1965,BrezinItzykson1970} [dashed line in Fig.~\ref{fig:concept}~(a)].
In the rest frame, the electric field transforms as
$E'_\parallel=E_\parallel,\; E'_\perp=E_\perp/\gamma_t$ 
for the components of the electric field parallel and perpendicular to the wavevector.  
Thus, the Keldysh parameter is reduced by $\gamma_t$ for a longitudinal wave near the shockwave limit,
driving the system into the nonadiabatic regime.
For a circularly polarized (rotating)
field, the tunneling trajectory is curved in parameter space and acquires
quantum-geometric corrections~\cite{Berry1990}; this regime was studied as the
twisted Schwinger effect in Ref.~\cite{Takayoshi2021} in the homogeneous
case. Floquet band renormalization
and carrier creation must therefore be treated together once the shockwave limit
drives the system into the nonadiabatic regime.

The aim of this study is to examine these shockwave-enhanced Floquet effects in
relativistic quasiparticles. We first formulate a Floquet--Bloch theory for
propagating waves in which the driving frequency and wavenumber are combined into a
single space--time reciprocal vector $q^\mu=(\omega,\bm{Q})$. Working in Sambe's
extended space--time picture~\cite{Sambe1973}, we show that replica bands are
generated along $q^\mu$ and that the quasienergy--momentum plane acquires a
space--time Brillouin zone. Systems with such intertwined space--time periodicity
have been discussed as space--time
crystals~\cite{XuWu2018,Peng2022,GaoNiu2022}, and propagating-wave driving has
recently been analyzed for acoustoelectric transport in
nanowires~\cite{yang2024quantizedacoustoelectricfloqueteffect} and for flat-band
engineering with surface polaritons~\cite{walicki2024floquetengineeringnearlyflat};
here we organize these ingredients around the Lorentz structure of the drive.
Within this framework we construct the geometric response, defining the Berry
connection and curvature of Floquet--Bloch states following the Bloch-state
formulation~\cite{RevModPhys.82.1959}, and extending the Floquet TKNN formula~\cite{OkaAoki2009} and the
polarization current to propagating-wave driving;
related Berry-phase quantities for space--time crystals have been obtained in a
semiclassical wave-packet framework~\cite{GaoNiu2022}. Equipping space--time with
the effective Minkowski metric set by $v_F$~\cite{Bjorken:100769} then yields a
Lorentz-invariant classification of the drive into time-like, light-like, and
space-like waves, together with the rest frames in which a time-like wave becomes
spatially uniform and a space-like wave becomes static. We apply this machinery to
two cases. For a two-dimensional Dirac fermion driven by an in-plane circular gauge
wave---a geometry relevant to strain, acoustic, polaritonic, and near-field
drives---the boost mixes the longitudinal vector potential with a scalar potential,
and we obtain a shockwave-enhanced Dirac mass and a controllable gap closing. For a
three-dimensional Dirac fermion driven by a transverse circular wave, we compute the
full Floquet--Bloch spectra and show how the same mechanism creates and tilts Weyl
nodes~\cite{PhysRevB.93.155107,Hubener2017,PhysRevB.96.041126}, connecting to the
light-induced chiral gauge fields observed
optically~\cite{yoshikawa2022lightinducedchiralgaugefield}. 
Our findings provide a new way to engineer band topology of quasiparticles in quantum materials using propagating fields such as laser light~\cite{Wang2013, McIver2024, Ito2023,yoshikawa2022lightinducedchiralgaugefield}, slow light~\cite{BabaSlowLight}, polaritons~\cite{Basov2021, walicki2024floquetengineeringnearlyflat}, acoustic waves~\cite{Oliva-Leyva_2016,Puebla_2020}, and sliding density waves~\cite{RevModPhys.60.1129, PhysRevLett.31.462, PhysRevLett.37.602, PhysRevLett.42.1498, Cox2008}, with potential experimental realizations in tunable Weyl semimetals.

\section{Floquet--Bloch theory for propagating waves}
\label{sec:formalism}

\subsection{Floquet--Bloch equation}
\label{sec:FB-equation}
We consider a non-interacting electron system described by a one-body wave function $\ket{\Psi(\bs{x},t)}$ satisfying 
the Schr\"odinger equation $i\pa_t\ket{\Psi}=(H_0(\bs{x})+V(\bs{x},t))\ket{\Psi}$. 
Here, $H_0(\bs{k})$ denotes the unperturbed Bloch Hamiltonian in momentum
space.  We set $\hbar=1$ throughout.
We study a propagating wave perturbation 
\begin{align}
 V(x,t)=V_1\cos(\bs{Q}\cdot \bs{x}-\omega t)+V_2\sin(\bs{Q}\cdot \bs{x}-\omega t),
 \label{eq:pot1}
\end{align} 
where $V_1$ and $V_2$ are space- and time-independent Hermitian matrices,
$V_j^\dagger=V_j$, acting on internal degrees of freedom such as spin,
sublattice, or orbital indices.  The wave has wavevector $\bs{Q}$, wave number $Q=|\bs{Q}|$, frequency $\omega$,
and speed $v=\omega/Q$, defining a {\it space-time reciprocal vector} given by 
$q^\mu=(\omega,\bs{Q})$.  
Below, we choose the $x$ axis along $\bs{Q}$.
Here on, we use the four-vector notation with 
the space-time and momentum-energy variables
$x^\mu=(t,\bs{x})$ and $k^\mu=(\ve,\bs{k})$,
and product $k\cdot x=k^\mu  x_\mu=\ve t-\bs{k}\cdot \bs{x}$.
We  introduce a Floquet-Bloch decomposition of the wave function as
\begin{align}
\ket{\Psi(\bs{x},t)}=\frac{1}{\sqrt{N}}e^{-ik\cdot x}\ket{\Phi(q\cdot x)}
=\frac{1}{\sqrt{N}}e^{-ik\cdot x}\sum_{n=-\infty}^\infty\ket{\Phi^n} e^{-inq\cdot x},
\label{eq:FBdecomposition}
\end{align}
where the Floquet-Bloch state $\ket{\Phi(q\cdot x)}$ has a periodicity of
$q\cdot x\sim q\cdot x+2\pi $ ($N$ is a normalization factor). 

\begin{figure}[tb]
\centering
\includegraphics[width=0.96\linewidth]{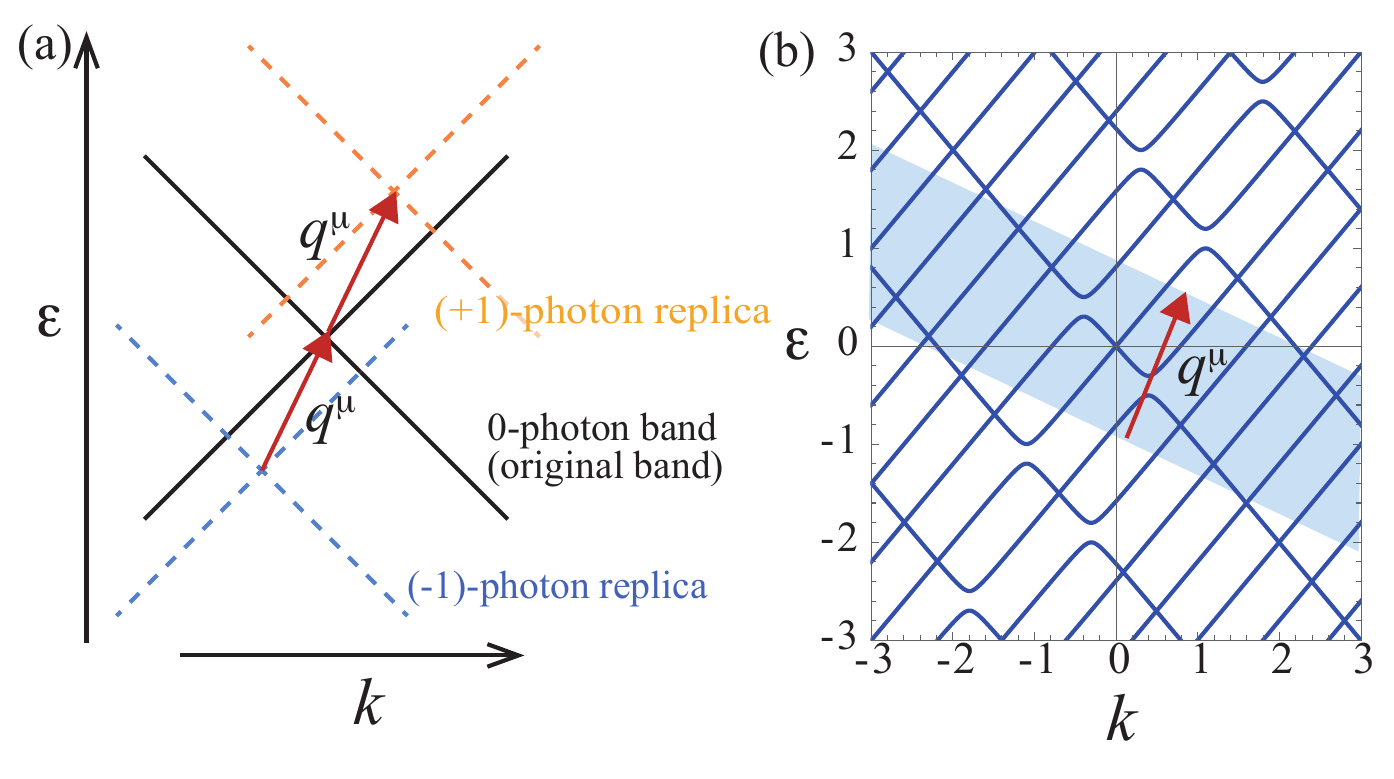}
\caption{
(a) 
Floquet--Bloch band folding along the space--time reciprocal pair
$(\omega,\bs{Q})$.
(b)
Computed Floquet--Bloch spectrum for a one-dimensional Dirac fermion with
$V_0=0.2$, $Q=0.7$, and $\omega=1.5$.  Quasienergy--momentum points related by
$(\Delta k_x,\Delta\ve)=(Q,\omega)$ belong to the same equivalence class.
The blue strip denotes one choice of the first space--time Brillouin zone,
centered at $(k,\varepsilon)=(0,0)$ and bounded by two lines perpendicular to
$q^\mu$; its opposite edges are identified by the simultaneous translation
$(\varepsilon,k)\to(\varepsilon+\omega,k+Q)$, indicated by the arrow.
}
\label{fig:schematic}
\end{figure}

We employ Sambe's space--time picture and write
$|\Phi_{\bs{k}}\dket=(\ldots,\ket{\Phi^{n-1}_{\bs{k}}},
\ket{\Phi^{n}_{\bs{k}}},\ket{\Phi^{n+1}_{\bs{k}}},\ldots)^{T}$
as an infinite dimensional vector with Fourier modes as components. 
Then, the time-dependent Schr\"odinger equation is recast as the
Floquet-Bloch equation (see also Ref.~\cite{yang2024quantizedacoustoelectricfloqueteffect}; the derivation, and the reason why $\ket{\Phi}$ depends on $x$ only through $q\cdot x$, are given in Appendix~\ref{app:FB})
\begin{align}
\mathcal{H}_{\bs{k}}|\Phi_{\alpha\bs{k}}\dket=\ve_{\alpha\bs{k}}|\Phi_{\alpha\bs{k}}\dket,
\label{eq:FBequation}
\end{align}
with the Floquet--Bloch Hamiltonian, where
$V=(V_1-iV_2)/2$ and $V^\dagger=(V_1+iV_2)/2$,
\begin{align}
\mathcal{H}_{\bs{k}}^{nn'}=
\left[H_0(\bs{k}+n\bs{Q})-n\omega\right]\delta_{nn'}
+V\delta_{n-1,n'}+V^\dagger\delta_{n+1,n'}.
\label{eq:FBequation2}
\end{align}
The Floquet--Bloch representation has a reciprocal-lattice redundancy:
shifting the quasienergy--momentum vector by an integer multiple of
$q^\mu$ and relabeling the Sambe index $n$ describes the same physical state,
\begin{align}
k^\mu\sim k^\mu+\ell q^\mu,\qquad \ell\in\mathbb{Z}.
\label{eq:FB-equivalence}
\end{align}
Systems with such intertwined space--time periodicity are known as
space--time crystals~\cite{XuWu2018,Peng2022,GaoNiu2022}; a single traveling wave
realizes the simplest ``oblique'' case, generated by one reciprocal vector $q^\mu$.
For a wave propagating along $x$, this identification reads
$(\ve,k_x)\sim(\ve+\ell\omega,k_x+\ell Q)$.  A first space--time
Brillouin zone is therefore a fundamental domain of this translation.  With
only one traveling wave, the zone is compact only along this reciprocal
translation and remains unbounded in the independent transverse direction in quasienergy--momentum space.
Any strip bounded by two parallel lines that are mapped onto each other by the translation~(\ref{eq:FB-equivalence}) is a valid choice.
Two natural ones are the strip perpendicular to $q^\mu$ in the Euclidean sense of the $(k_x,\ve)$ plane, used in Fig.~\ref{fig:schematic}~(b),
and the Lorentz-covariant strip $|q\cdot k|\le|q\cdot q|/2$ [with the four-vectors scaled by $v_F$ as in Sec.~\ref{sec:Lorentz-framework}],
which is the image of the rest-frame Floquet zone $|\ve'|\le\omega'/2$ (time-like) or Bloch zone $|k'_x|\le Q'/2$ (space-like) under the boost and is used in Fig.~\ref{fig:Weylpoints}.
The width of the latter, proportional to $|\omega^2-v_F^2Q^2|$, vanishes in the light-like limit, where the replica bands fold infinitely densely.

\textit{Demonstration with a 1D model.}---
To illustrate these structures, we consider the
minimal one-dimensional Dirac Hamiltonian
\begin{align}
H_0(k_x)=v_F\sigma_x k_x
\label{eq:H1D}
\end{align}
driven by a single wave propagating in the $x$-direction
\begin{align}
V(x,t)=V_0\sigma_y\cos(Qx-\omega t).
\label{eq:V1D}
\end{align}
As depicted in Fig.~\ref{fig:schematic}~(a), in the absence of coupling between the Sambe sectors ($V_0=0$), the spectrum 
of Eq.~(\ref{eq:FBequation}) consists
of replicas (known as photon dressed replicas)
\begin{align}
\ve^{(0)}_{s,n}(k_x)=s v_F(k_x+nQ)-n\omega,
\qquad s=\pm,
\label{eq:replica1D}
\end{align}
obtained by translating the right- and left-moving Dirac branches
simultaneously in momentum and energy.  The perturbation in
Eq.~(\ref{eq:V1D}) connects neighboring sectors,
$n\leftrightarrow n\pm1$.  Since $\sigma_y$ is off diagonal in the eigenbasis
of $\sigma_x$, it hybridizes counterpropagating branches with opposite
$\sigma_x$ eigenvalues.  Their intersections consequently become avoided
crossings, producing the gaps visible in Fig.~\ref{fig:schematic}~(b), where
we set $v_F=1$, $V_0=0.2$, $Q=0.7$, and $\omega=1.5$.
The blue strip in Fig.~\ref{fig:schematic}~(b) is one convenient
choice: every equivalence class has one representative in the strip, and
points on its two edges are identified by the same translation.

This example interpolates continuously between the conventional Floquet
problem at $Q=0$, where folding occurs only in quasienergy $\ve\sim \ve+\ell \omega$, and a static Bloch
superlattice at $\omega=0$, where folding occurs only in momentum $k\sim k+\ell Q$.  For finite
$Q$ and $\omega$, energy and momentum are transferred together along
$(\omega,Q)$.  The resulting tilted replica structure and avoided
crossings provide the simplest illustration of the Floquet--Bloch
construction used below for two- and three-dimensional Dirac fermions.

We stress that the Floquet--Bloch construction itself is not restricted to relativistic quasiparticles.
For a non-relativistic band, $H_0(\bs{k})=\bs{k}^2/2m^*$, a Galilean transformation to the frame co-moving with the wave,
$\Psi(\bs{x},t)=e^{i(m^*vx-m^*v^2t/2)}\Psi'(x-vt,y,z,t)$, removes the time dependence of Eq.~(\ref{eq:pot1}) for {\it any} wave speed,
and the problem reduces to the static Bloch problem of a particle with momentum $k_x-m^*v$ in the periodic potential;
the quasienergy is $\ve=\ve'(k_x-m^*v)+v(k_x-m^*v)+m^*v^2/2$, a mere tilt of the static minibands.
There is no distinguished wave speed and no enhancement.
The classification into time-, light-, and space-like waves and the Lorentz factors of the following sections are therefore specific to the linear (relativistic) dispersion;
for a lattice band with bounded velocity, the relevant counterpart is the matching of $v$ with the group velocity,
where the replica bands become degenerate along a line in $k$ space (the acoustoelectric resonance of Ref.~\cite{yang2024quantizedacoustoelectricfloqueteffect}).

\subsection{Geometric response}
\label{sec:geometry}
The formulation of the geometric properties of the Floquet--Bloch states follows
those of the Bloch states~\cite{RevModPhys.82.1959}. Berry-phase quantities for
space--time crystals, including the polarization of a filled Floquet--Bloch band
and the resulting current response, have been developed within a semiclassical
wave-packet framework in Ref.~\cite{GaoNiu2022}; here we organize them in a form
adapted to the Lorentz classification used below.

We define the Berry connection of the $\alpha$-th state by
\begin{align}
&\mathcal{A}_{\alpha\bs{k},a}(s)
=\bra{\Phi_{\alpha\bs{k}}(s)}i\pa_{k_a}
\ket{\Phi_{\alpha\bs{k}}(s)},\qquad s=\bs{Q}\cdot\bs{x}-\omega t=-q\cdot x.
\label{eq:BerryA}
\end{align}
The Berry curvature of the Floquet-Bloch state can be defined by~\cite{RevModPhys.82.1959}
\begin{align}
\Omega_{\alpha\bs{k},ab}(s)
=\pa_{k_a}\mathcal{A}_{\alpha\bs{k},b}(s)
-\pa_{k_b}\mathcal{A}_{\alpha\bs{k},a}(s),
\label{eq:BerryOmega}
\end{align}
which is gauge invariant. 
We can formulate the 
Floquet TKNN formula~\cite{OkaAoki2009}  as
\begin{align}
\sigma_{xy}=e^2\int\frac{d\bs{k}}{(2\pi)^d}\sum_{\alpha}f_{\alpha}(\bs{k})
\int_0^{2\pi} \frac{ds}{2\pi}
\Omega_{\alpha\bs{k},xy}(s)
\label{eq:TKNN}
\end{align}
using the distribution function $f_{\alpha}(\bs{k})$~\cite{SchulerPRX2020,OkaAoki2009,Dehghani2014,Dehghani2015,Genske2015,Seetharam2019}, 
where $d$ is the spatial dimension.
This result extends the Chern number formulation to propagating-wave-driven systems.
 Floquet-Bloch states also exhibit non-adiabatic polarization current~\cite{Resta1992,King-Smith93} and geometric phase effects~\cite{Berry1984}. 
Integrating the Berry connection along the wave-vector direction, we define
\begin{align}
&P^x_{\alpha\bs{k}_\perp}(s)=\frac{1}{2\pi}\int_{0}^{Q}dk_x \bra{\Phi_{\alpha\bs{k}}(s)} (i\pa_{k_x})\ket{\Phi_{\alpha\bs{k}}(s)},
\label{eq:Pk}\\
&\gamma_{\alpha\bs{k}}=\int_0^{2\pi}ds\,\bra{\Phi_{\alpha\bs{k}}(s)} (i\pa_{s})\ket{\Phi_{\alpha\bs{k}}(s)},
\label{eq:Berryphase}
\end{align}
whose gauge invariance (modulo an integer and $2\pi$, respectively) follows from the periodicity of the Floquet-Bloch state. 
Here $\gamma_{\alpha\bs{k}}$ is the Berry phase along the phase variable $s$; since $s=-\omega t$ at fixed $\bs{x}$, the Berry phase accumulated over one period in time, $i\oint dt\,\bra{\Phi}\pa_t\ket{\Phi}$, is $-\gamma_{\alpha\bs{k}}$.

Here, $\bs{k}_\perp=(0,k_y,k_z)$ is the crystal momentum 
perpendicular to the wave propagation direction $\bs{Q}$. 
The geometric phase $\gamma_{\alpha\bs{k}}$ is well studied in Floquet systems
and is related to topological pumping (see, e.g., Ref.~\cite{Nakagawa2020}).
Using Eq.~(\ref{eq:Pk}),
we can define the total polarization by 
\begin{align}
P^x_{\rm tot}(Qx-\omega t)=
e\int\frac{d\bs{k}_\perp}{(2\pi)^{d-1}}\sum_{\alpha}f_{\alpha}(\bs{k}_\perp)
P^x_{\alpha\bs{k}_\perp}(Qx-\omega t),
\label{eq:Px}
\end{align}
where each band is assumed to be uniformly occupied along the propagation direction, $f_\alpha(\bs{k}_\perp)$, as required for the gauge invariance of the $k_x$ integral in Eq.~(\ref{eq:Pk}),
whose derivative gives the longitudinal polarization current $J^x_{\rm tot}(Qx-\omega t)=\frac{d}{dt}
P^x_{\rm tot}(Qx-\omega t)$ flowing parallel to the propagating wave.

\begin{figure}[tb]
\centering
\includegraphics[width=0.6\linewidth]{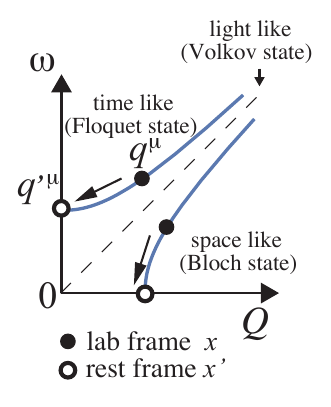}
\caption{
Lorentz transformation of the space--time reciprocal vector. A Lorentz transformation maps propagating waves into a ``rest frame,''
where space-like (time-like) waves appear static (homogeneous).
The solid lines ($\omega^2/v_F^2 - Q^2 = \text{const.}$) represent the
Lorentz orbits of the scaled reciprocal vectors
$q^\mu=(\omega/v_F,\bs{Q})$.
Filled (open) circles mark $q^\mu$ in the laboratory (rest) frame: a time-like vector is boosted onto the $\omega$ axis, where the rest-frame problem is a Floquet problem with the reduced frequency $\omega/\gamma_t$,
and a space-like vector onto the $Q$ axis, where it is a Bloch problem with the reduced wave number $Q/\gamma_s$; a light-like vector (dashed line) cannot be brought to rest and corresponds to the Volkov state.
}
\label{fig:Lorentz_momentum}
\end{figure}

\subsection{Lorentz classification and rest frames}
\label{sec:Lorentz-framework}
We now specialize the Floquet--Bloch formalism to relativistic quasiparticles. 
To make the argument concrete, we consider the three-dimensional Dirac model, which will be studied further in
Sec.~\ref{sec:3D}.  The focus here is to establish how the Floquet-Bloch states transform under 
Lorentz transformation and set the classification scheme. 

The Hamiltonian can be written as (see, e.g., Ref.~\cite{Bjorken:100769})
\begin{align}
H_0(\bs{x})=v_F\sum_i\Gamma^i(\hat{p}_i+eA_i+eA_{5i}\gamma^5)+m\gamma^0-eA_0I-eA_{0}^5\gamma^5,
\label{eq:HamDirac}
\end{align}
with gamma matrices $\gamma^0=
\left(
\begin{array}{cc}
  0&I_2      \\
  I_2& 0    
\end{array}
\right),\;\Gamma^i=
\left(
\begin{array}{cc}
  -\sigma_i&0      \\
  0& \sigma_i
\end{array}
\right)
$ 
and $\gamma^5=\left(
\begin{array}{cc}
  -I_2&0      \\
  0& I_2
\end{array}
\right)$. 
Here,  $m$ is the mass parameter, $A_\mu$ denotes external electromagnetic fields 
(the electron charge is $-e$, and $(A_0\/v_F,\bs{A})$ transforms as a Lorentz four-vector), 
and 
 $A_{5\mu}$ denotes a chiral gauge field that can separate a Dirac node into
Weyl nodes~\cite{YanFelser2017}.  The Dirac equation is Lorentz covariant, with
the Fermi velocity $v_F$ playing the role of the speed of light.  

Accordingly, we now equip space--time with the effective Minkowski line element
$d\mathsf{s}^2=v_F^2dt^2-d\bs{x}^2$.  Equivalently, we rescale the time components of the four-vectors of Sec.~\ref{sec:FB-equation} by $v_F$, so that all components have common units,
\begin{align}
x^\mu=(v_Ft,\bs{x}),\qquad
k^\mu=(\ve/v_F,\bs{k}),\qquad
q^\mu=(\omega/v_F,\bs{Q}),
\label{eq:scaled-four-vectors}
\end{align}
and the Minkowski metric $\eta_{\mu\nu}=\mathrm{diag}(1,-1,-1,-1)$.
The phases introduced in Sec.~\ref{sec:FB-equation} are unchanged, $k\cdot x=\ve t-\bs{k}\cdot\bs{x}$ and $q\cdot x=\omega t-\bs{Q}\cdot\bs{x}$, so that all results of Secs.~\ref{sec:FB-equation} and \ref{sec:geometry} carry over and we keep the same symbols.
Unlike the general covectors used there, however, the scaled
$k^\mu$ and $q^\mu$ transform as Lorentz four-vectors.  In
particular,
\begin{align}
q^2=\frac{\omega^2}{v_F^2}-\bs{Q}^2
\label{eq:q-invariant}
\end{align}
provides an invariant classification of the wave.
We classify the waves as:
\begin{itemize}
\item Time-like ($q^2>0$, i.e., $v > v_F$): Wave moves faster than the Fermi velocity; a boost makes it spatially homogeneous.
\item Light-like ($q^2=0$, $v = v_F$): Wave propagates at the Fermi velocity; no rest frame exists.
\item Space-like ($q^2<0$, $v < v_F$): Wave moves slower than the Fermi velocity; a boost makes it static.
\end{itemize}
To simplify the analysis, we transform into a rest frame in which the wave
becomes static (space-like) or spatially homogeneous (time-like), as shown in
Fig.~\ref{fig:Lorentz_momentum}.  A boost of velocity $u$ along $x$ acts as
\begin{align}
\begin{pmatrix}v_Ft'\\x'\end{pmatrix}
=
\begin{pmatrix}\gamma_u&-\gamma_u u/v_F\\
-\gamma_u u/v_F&\gamma_u\end{pmatrix}
\begin{pmatrix}v_Ft\\x\end{pmatrix},
\qquad
\gamma_u=\frac{1}{\sqrt{1-u^2/v_F^2}},
\label{eq:Lorentz-boost}
\end{align}
with $y'=y$ and $z'=z$.
For a time-like wave we choose a boost velocity $u=v_F^2/v$ and
$\gamma_t=(1-v_F^2/v^2)^{-1/2}$; for a space-like wave we choose $u=v$ and
$\gamma_s=(1-v^2/v_F^2)^{-1/2}$.  We then obtain the scaled reciprocal
vectors
\begin{align}
&q'{}^\mu=(\omega/(v_F\gamma_t),0,0,0)
\quad\mbox{(time-like)},
\label{eq:restframe1}
\\
&q'{}^\mu=(0,Q',0,0)
\qquad\mbox{(space-like)},
\label{eq:restframe2}
\end{align}
with the rest-frame wave number $Q'=Q/\gamma_s$.
Thus, a time-like wave becomes a spatially uniform oscillation, whereas a
space-like wave becomes a static spatial modulation.  The time-like
rest-frame frequency $\omega'=\omega/\gamma_t$ vanishes on approaching the
light line from above.
The Floquet--Bloch solutions in the laboratory and rest frames are related by
\begin{align}
\ket{\Psi_{\alpha\bs{k}}(x)}
={}&D(a^{-1})\ket{\Psi'_{\alpha\bs{k}'}(ax)}
\nonumber\\
={}&\frac{e^{-i(a^{-1}k')\cdot x}}{\sqrt{N}}D(a^{-1})
\ket{\Phi'_{\alpha\bs{k}'}(q\cdot x)},
\label{eq:LorentzState}
\end{align}
where $D(a)=e^{-\theta\Gamma^x/2}=\cosh(\theta/2)-\sinh(\theta/2)\Gamma^x$, with $\tanh\theta=u/v_F$, is the spinor representation of the boost,
which satisfies $D(a)^{-1}\gamma^\mu D(a)=a^\mu{}_\nu\gamma^\nu$ with $\gamma^i\equiv\gamma^0\Gamma^i$ (Appendix~\ref{app:Lorentz}).
The state $\ket{\Psi'(x')}$ is determined by solving the Floquet--Bloch equation~(\ref{eq:FBequation})
using fields in the rest frame
\begin{align}
A'{}^{\mu}&=a^{\mu}{}_\nu A^{\nu},\qquad
A_5'{}^{\mu}=a^{\mu}{}_\nu A_5{}^{\nu},
\nonumber\\
F'^{\mu\nu}&=a^{\mu}{}_\rho a^{\nu}{}_\sigma F^{\rho\sigma}.
\label{eq:Lorentzfield}
\end{align}
In general the Fourier matrix $V$ is not boost invariant.  Longitudinal
vector-potential components mix with the scalar potential, while transverse
components are unchanged.  This distinction is central to the two-dimensional
application in Sec.~\ref{sec:2D}.  The transverse circular drive used in
Sec.~\ref{sec:3D}, by contrast, is invariant under an $x$-directed boost at the
level of its vector-potential amplitudes.
From Eq.~(\ref{eq:LorentzState}), we
find that the scaled quasienergy--momentum four-vector
and the Floquet-Bloch state
in the two frames are related by 
$
k=a^{-1}k'$
and 
$\ket{\Phi(q\cdot x)}
=D(a^{-1})\ket{\Phi'(q\cdot x)}$.
The former relation implies that the spectra 
in the two frames are related by Lorentz boost
(See Fig.~\ref{fig:Weylpoints} below for an example).

\section{Application to two-dimensional Dirac fermions}
\label{sec:2D}

\begin{figure*}[tb]
\centering
\includegraphics[width=\linewidth]{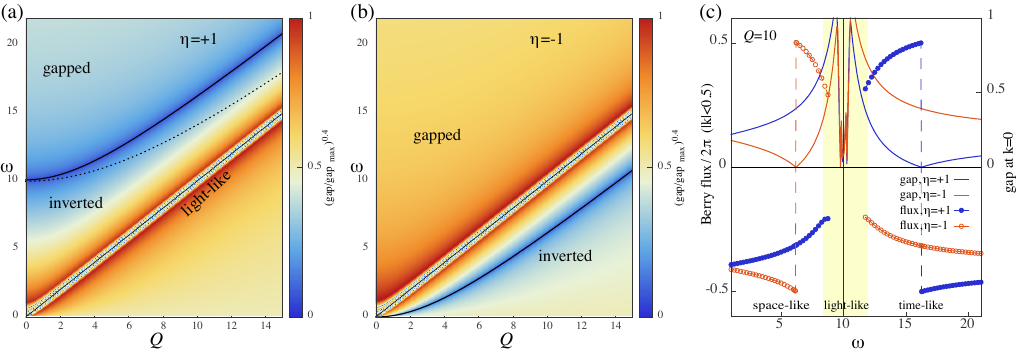}
\caption{
Two-dimensional massive Dirac fermion ($m=0.1$, $g=1$, $v_F=1$) in the circular gauge wave of Eq.~(\ref{eq:A2D}).
(a,b) Quasienergy gap at $\bs{k}=0$ in the $(Q,\omega)$ plane for helicity $\eta=+1$ (a) and $\eta=-1$ (b);
the color scale is $(\Delta/\Delta_{\max})^{0.4}$. Thin line: light-like line $\omega=v_FQ$.
Solid line in (a): time-like gap-closing line $\omega-Q^2/\omega=(g^2+m^2)/m$ [Eq.~(\ref{eq:2D-closing-numerical})];
dotted line: $\omega^2-Q^2=(g^2/m)^2$, the closing line that would follow from the frequency redshift alone.
Solid line in (b): space-like gap-closing line $(v/v_F)/(1-v^2/v_F^2)=mv_FQ/g^2$ [Eq.~(\ref{eq:spacelike-closing})].
(c) Gap at $\bs{k}=0$ for $Q=10$ and both helicities (right axis), 
and the Berry flux of the lower Dirac band through the disc $|\bs{k}|<0.5$ 
(left axis, in units of $2\pi$). 
The flux jumps by one at $\omega_c\simeq16.24$ for $\eta=+1$ and at $\omega_c\simeq6.18$ for $\eta=-1$. 
In the shaded window around the light line the lower band cannot be identified by its static weight and the flux is not shown.
}
\label{fig:dirac2d}
\end{figure*}

\subsection{Circular gauge wave and Floquet--Bloch Hamiltonian}

We consider a massive two-dimensional Dirac fermion
\begin{align}
H_{\rm 2D}^{(0)}(\bs{k})
=v_F(\sigma_x k_x+\sigma_y k_y)+m\sigma_z
\label{eq:H2D}
\end{align}
coupled by minimal coupling to an in-plane rotating vector potential,
\begin{align}
\bs{A}(x,t)&=A\left[\bs{e}_x\cos\phi
-\eta\bs{e}_y\sin\phi\right],
&\phi&=Qx-\omega t,
\label{eq:A2D}
\end{align}
where $\eta=\pm1$ is the helicity.  The driven Hamiltonian is
\begin{align}
H_{\rm 2D}(x,t)=v_F\bs{\sigma}\cdot
\left[\hat{\bs{p}}+e\bs{A}(x,t)\right]+m\sigma_z.
\label{eq:H2D-driven}
\end{align}
It is useful to define $g=e v_FA$.  The positive-phase Fourier matrix in the
convention of Eq.~(\ref{eq:FBequation2}) is
\begin{align}
V_{\rm 2D}=\frac{g}{2}(\sigma_x+i\eta\sigma_y).
\label{eq:V2D}
\end{align}
Consequently, the exact Sambe Hamiltonian is block tridiagonal,
\begin{align}
\mathcal{H}_{{\rm 2D},\bs{k}}^{nn'}={}&
\left\{v_F\left[\sigma_x(k_x+nQ)+\sigma_y k_y\right]
+m\sigma_z-n\omega\right\}\delta_{nn'}
\nonumber\\
&+V_{\rm 2D}\delta_{n-1,n'}
+V_{\rm 2D}^{\dagger}\delta_{n+1,n'}.
\label{eq:Sambe2D}
\end{align}
Thus, a circular gauge wave produces helicity-selective pseudospin flips
between neighboring Floquet--Bloch replicas.  Equation~(\ref{eq:Sambe2D}) is
also the appropriate starting point at the light line and in the
nonperturbative shockwave regime, where a truncated high-frequency expansion
is not controlled.

We emphasize the physical meaning of Eq.~(\ref{eq:A2D}).  
The homogeneous version of the model with $Q=0$ has been studied in the 
context of a Floquet Chern insulator~\cite{OkaAoki2009}. 
However, for a freely
propagating electromagnetic wave whose wave vector lies in the material
plane, both transverse polarizations cannot lie in that plane.  We therefore
refer to Eq.~(\ref{eq:A2D}) as a \emph{circular gauge wave}.  It can describe,
for example, a rotating strain-induced pseudogauge field, a coherent acoustic
or phonon mode, a polaritonic near field, or an engineered modulation.  It is
not assumed to be a transverse plane wave in vacuum.

\subsection{Rest-frame drive and enhanced Dirac mass}

Care is needed for the longitudinal field $A_x$ in Eq.~(\ref{eq:A2D}) since the 
Lorentz boost will change its value,
in contrast to the transverse fields considered later in Section~\ref{sec:3D}.
Introduce the electromagnetic four-potential as
$A^\mu=(A^0,A_x,A_y)$, with $A^0$ the scalar potential divided by $v_F$ [cf. Eq.~(\ref{eq:HamDirac})], and take
$A^0=0$ in the laboratory frame.  For a
time-like wave, let
\begin{align}
\beta=\frac{v_F}{v},\qquad
\gamma_t=\frac{1}{\sqrt{1-\beta^2}},\qquad
\omega'=\frac{\omega}{\gamma_t}.
\label{eq:2D-boost-parameters}
\end{align}
The boost with velocity $u=v_F^2/v$ then gives
\begin{align}
A'^0(t')&=-\gamma_t\beta A\cos\phi',
&A_x'(t')&=\gamma_t A\cos\phi',
\nonumber\\
A_y'(t')&=-\eta A\sin\phi',\qquad \phi'=-\omega't'.
\label{eq:2D-boosted-potential}
\end{align}
In particular, setting the transformed scalar potential to zero would violate
Lorentz covariance.  Combining its identity-matrix coupling $-ev_FA'^0I$ [Eq.~(\ref{eq:HamDirac})] 
with the vector coupling, the rest-frame Fourier matrix becomes
\begin{align}
V'_{\rm 2D}=\frac{g}{2}
\left[\gamma_t(\sigma_x+\beta I)+i\eta\sigma_y\right].
\label{eq:V2D-rest}
\end{align}

When $\lVert V'_{\rm 2D}\rVert/\omega'\ll1$, a van Vleck expansion gives~\cite{PhysRevB.84.235108,GoldmanDalibard2014}
\begin{align}
H'_{\rm eff}
&=H_{\rm 2D}^{(0)}+\frac{[V_{\rm 2D}'{}^\dagger,V'_{\rm 2D}]}{\omega'}
+O(\omega'^{-2})
\nonumber\\
&=v_F(\sigma_xk_x'+\sigma_yk_y')+m_{\rm eff}'\sigma_z
+O(\omega'^{-2}),
\label{eq:Heff2D}
\\
m_{\rm eff}'
&=m-\eta\frac{\gamma_t^2g^2}{\omega}.
\label{eq:meff2D}
\end{align}
Since $V'_{\rm 2D}$ couples the $n$-th Floquet replica to the $(n-1)$-th one, Eq.~(\ref{eq:Heff2D}) is Eq.~(\ref{eq:1st}) with $H_{+1}=V'_{\rm 2D}$ and $\omega\to\omega'$; for $\eta=+1$ and $Q=0$ it agrees with the exact solution in Appendix~\ref{app:exact}.
The identity component is required to maintain Lorentz covariance.
In the spatially homogeneous time-like rest frame, however, 
it is a periodic scalar potential with zero mean and can be removed by a periodic time-dependent phase,
so that it does not affect the quasienergy spectrum.

The factor $\gamma_t^2$ in Eq.~(\ref{eq:meff2D}) originates from the boost enhancement of the longitudinal vector-potential amplitude combined with the frequency redshift.  
Reversing the helicity reverses the induced
mass.  The rest-frame gap therefore closes, within the controlled regime of
the expansion, when
\begin{align}
m=\eta\frac{\gamma_t^2g^2}{\omega}.
\label{eq:2D-gap-closing}
\end{align}
Closer to $v=v_F$, Eq.~(\ref{eq:meff2D}) must not be extrapolated to a
divergence: both $\omega'$ decreases and $V'_{\rm 2D}$ grows, so the expansion
eventually fails.  The full Floquet--Bloch matrix~(\ref{eq:Sambe2D}), or its
Lorentz-transformed counterpart including Eq.~(\ref{eq:V2D-rest}), then
describes the crossover to the nonperturbative shockwave state.

{\it Space-like waves.}---The same mechanism operates on the Bloch side of the light line, with the opposite helicity.
For a space-like wave the rest frame moves with $u=v$, $\beta_s=v/v_F$, and the boosted drive is static,
$V'_{\rm 2D}=\frac{g}{2}[\gamma_s(\sigma_x+\beta_sI)+i\eta\sigma_y]$ with wave number $Q'=Q/\gamma_s$.
Second-order perturbation theory in the static superlattice, in which the state at $\bs{k}'=0$ couples to the Bloch copies at $k'_x=\pm Q'$ with Hamiltonians $\pm v_FQ'\sigma_x+m\sigma_z$, gives (Appendix~\ref{app:exact})
\begin{align}
m_{\rm eff}'=m+g^2\,\frac{m+\eta\,\gamma_s\beta_s\,v_FQ'}{v_F^2Q'^2+m^2}.
\label{eq:meff-spacelike}
\end{align}
The helicity-independent part merely enhances $|m|$, whereas the helicity-dependent part,
which arises from the cross term between the scalar potential generated by the boost and the transverse vector potential,
reverses the sign of the mass for $\eta=-1$ when
\begin{align}
\frac{\gamma_s^2\beta_s\,g^2}{v_FQ}=m,\qquad{\rm i.e.}\qquad \frac{v/v_F}{1-v^2/v_F^2}=\frac{mv_FQ}{g^2}.
\label{eq:spacelike-closing}
\end{align}
Thus the two helicities invert the Dirac mass on opposite sides of the light line: $\eta=+1$ in the Floquet regime [Eq.~(\ref{eq:2D-gap-closing})] and $\eta=-1$ in the Bloch regime.
Equation~(\ref{eq:spacelike-closing}) is the solid line in Fig.~\ref{fig:dirac2d}~(b) (see Sec.~\ref{sec:2D-numerics}); for $m=0.1$, $g=1$ it predicts $\omega_c=6.18$ ($2.07$) for $Q=10$ ($5$),
in agreement with the numerical gap closing between $\omega=5.8$ and $6.4$ ($2.0$ and $2.3$), across which $\langle\sigma_z\rangle$ of the lower state at $\bs{k}=0$ changes sign
and the Berry flux jumps by one [Fig.~\ref{fig:dirac2d}~(c)]. In both regimes the enhancement near the light line ($\gamma_{t,s}\to\infty$) is driven by the longitudinal component of the gauge wave.

\subsection{Numerical results}
\label{sec:2D-numerics}
In order to understand how the topological transition occurs, 
we plot the quasienergy gap at $\bs{k}=0$ in the $(Q,\omega)$ plane for both helicities in Fig.~\ref{fig:dirac2d}~(a,b).
This is done by diagonalizing Eq.~(\ref{eq:Sambe2D}) with $2N+1=25$ replicas for $m=0.1$ and $g=1$ (units $v_F=1$).
For $\eta=+1$ [Fig.~\ref{fig:dirac2d}~(a)] the gap closes along a line in the $(Q,\omega)$ plane in the time-like regime and reopens with the inverted mass below it.
This is in contrast to the case of $\eta=-1$ [Fig.~\ref{fig:dirac2d}~(b)], where it does not close in the time-like regime, 
but inverts in the space-like regime along the solid line of Eq.~(\ref{eq:spacelike-closing}).
Figure~\ref{fig:dirac2d}~(c) shows, for $Q=10$, the gap for both helicities together with the Berry flux of the lower Dirac band
[the Floquet--Bloch state continuously connected to the undriven valence band, selected by its static weight $|\Phi^0_{\alpha\bs{k}}|^2$]
through a disc $|\bs{k}|<K$ around the node,
$\int_{|\bs{k}|<K}\frac{d\bs{k}}{2\pi}\int_0^{2\pi}\frac{ds}{2\pi}\,\Omega_{\alpha\bs{k},xy}(s)$, i.e., the contribution of the disc to the Floquet TKNN formula~(\ref{eq:TKNN}) in units of $e^2/h$.
The flux jumps from $-1/2$ to $+1/2$ at $\omega_c\simeq16.24$ for $\eta=+1$, while for $\eta=-1$ the jump occurs on the space-like side at $\omega_c\simeq6.18$ [Eq.~(\ref{eq:spacelike-closing})]; in each case the Hall conductance of the cone changes by $e^2/h$ across the transition,
which identifies the gap closing as a change of the Floquet Chern number; the absolute value is subject to the regularization discussed below.

At $Q=0$ the problem is exactly solvable at $\bs{k}=0$ (rotating frame, cf.\ Appendix~\ref{app:exact}),
$\ve=\pm\sqrt{g^2+(m-\eta\omega/2)^2}+\omega/2$, so that the gap closes at $\omega_c=(g^2+m^2)/m$;
for $Q\ne0$ we find numerically that the closing line is described 
to within $0.1\%$ (for $m\ll g$) up to $Q=15$ by
\begin{align}
\frac{\omega}{\gamma_t^2}=\omega-\frac{Q^2}{\omega}=\frac{g^2+m^2}{m},
\label{eq:2D-closing-numerical}
\end{align}
i.e., Eq.~(\ref{eq:2D-gap-closing}) with $g^2\to g^2+m^2$ [solid line in Fig.~\ref{fig:dirac2d}~(a)].
The $\gamma_t^2$ dependence is clearly distinguished from the hyperbola $\omega^2-Q^2=(g^2/m)^2$ (dotted line)
that would follow from the frequency redshift alone. The closing occurs at $\bs{k}=0$: a scan over $|\bs{k}|<3$ finds no smaller gap elsewhere.

Close to the light line the gap at $\bs{k}=0$ collapses and oscillates on a fine scale [streaks parallel to $\omega=v_FQ$ in Figs.~\ref{fig:dirac2d}~(a,b)],
reflecting the dense folding of the replicas in the shockwave regime: 
the Floquet--Bloch state at $\bs{k}=0$ spreads over a number of replicas that grows as $\omega'\to0$.
This signals the breakdown of the few-photon (perturbative) picture, beyond which nonperturbative processes such as tunneling can set in.
Exactly at the light line $q^2=0$, no rest frame exists, 
the Lorentz-covariant Floquet--Bloch zone inherited from the rest frame collapses, and the replica participation is expected to diverge in the continuum Dirac model;
this shockwave state separates the Floquet (time-like) and Bloch (space-like) regimes.

\begin{figure}[tb]
\centering
\includegraphics[width=\linewidth]{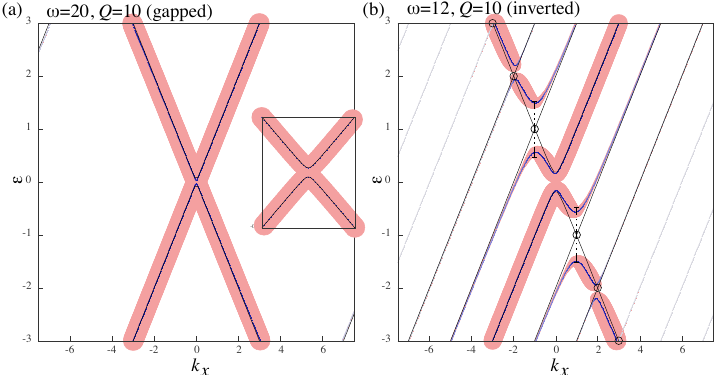}
\caption{
Floquet--Bloch quasienergy spectra along $k_y=0$ of the two-dimensional model of Fig.~\ref{fig:dirac2d} ($m=0.1$, $g=1$, $v_F=1$, $\eta=+1$) at $Q=10$,
for (a) $\omega=20$ (time-like, above the gap-closing line) and (b) $\omega=12$ (time-like, below it).
The size of the red discs and the intensity of the blue dots give the static weight $\langle\Phi^0_{\alpha\bs{k}}|\Phi^0_{\alpha\bs{k}}\rangle$ of each state.
Black dashed lines: the analytic high-frequency result, the dressed Dirac cone of Eq.~(\ref{eq:Heff2D}) boosted back to the laboratory frame,
$(\ve-n\omega)^2-v_F^2(k_x-nQ)^2=m_{\rm eff}'^2$ with $m'_{\rm eff}=m-\gamma_t^2g^2/\omega$ and its replicas $n=0,\pm1,\pm2$.
The inset in (a) magnifies the gap at $\bs{k}=0$ ($2|m'_{\rm eff}|=0.067$).
In (b) the mass is inverted ($m'_{\rm eff}=-0.17$, gap $0.35$) and the spectrum shows the hybridization with replica copies.
}
\label{fig:dirac2d-spectra}
\end{figure}

Figure~\ref{fig:dirac2d-spectra} compares the quasienergy spectra along $k_y=0$ at $Q=10$ with the analytic high-frequency result.
Since Eq.~(\ref{eq:Heff2D}) is a Dirac Hamiltonian in the rest frame, its spectrum is Lorentz covariant and appears in the laboratory frame simply as the hyperbola
$\ve^2-v_F^2k^2=m_{\rm eff}'^2$, together with its replicas shifted by $(nQ,n\omega)$; 
this reproduces the dressed cone and the gap at $\bs{k}=0$ both above ($\omega=20$) and below ($\omega=12$) the closing line.

Beyond the effective Hamiltonian, 
we see gaps opening at the $n$-photon resonances where
the replica cones cross the original cone on the branch counter-propagating to the wave; 
in the rest frame these are the ordinary multiphoton resonances $2\sqrt{v_F^2k'^2+m^2}=n\omega'$,
and the one-photon gap follows from the matrix element of $V'_{\rm 2D}$ between the Dirac bands, in particular, we have $\Delta_1=g\,|\eta+2\gamma_t^2m/\omega|$ for the one-photon gap.
The spacing of the resonances vanishes at the light line, where the folding becomes dense.


\begin{figure*}[tb]
\centering
\includegraphics[width=\linewidth]{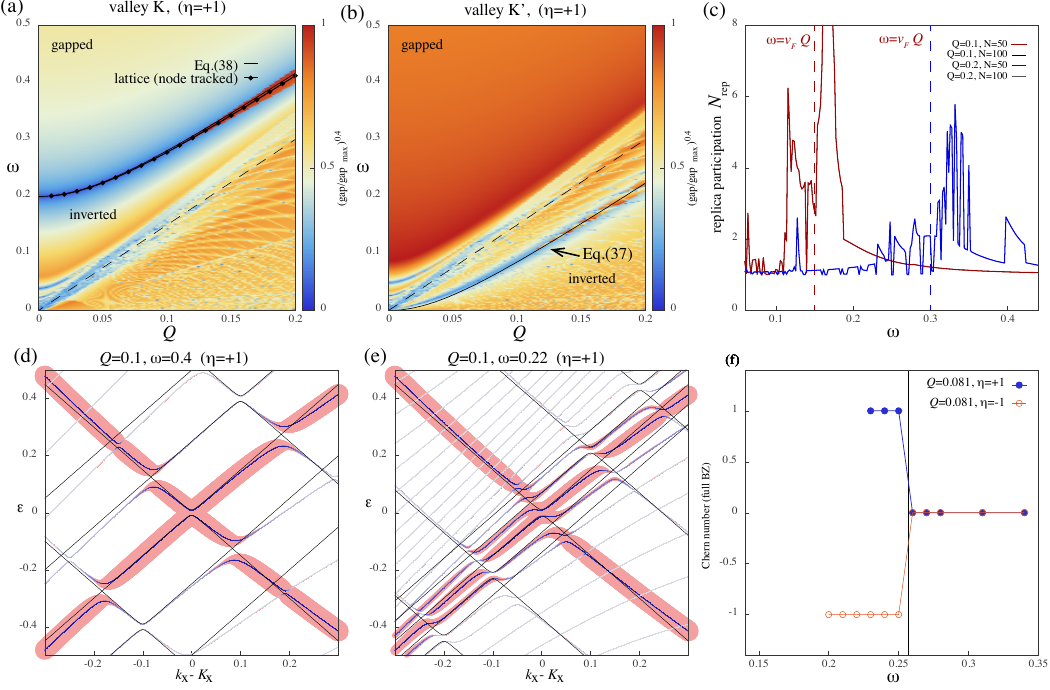}
\caption{
Honeycomb lattice with a staggered potential ($t=a=1$, $v_F=1.5$, $m=0.02$, $A=0.04$, $g=0.060$) driven by the circular gauge wave along $x$.
(a,b) Gap at $\bs{K}$ and at $\bs{K}'$ for $\eta=+1$ [color scale $(\Delta/\Delta_{\max})^{0.4}$]. Thin dashed line: $\omega=v_FQ$; thin solid lines: continuum closing lines, Eqs.~(\ref{eq:2D-closing-numerical}) and (\ref{eq:spacelike-closing});
dots in (a): closing line of the lattice model obtained by tracking the node.
(c) Replica participation number of the state closest to $\ve=0$ at $\bs{K}$ for $Q=0.1$ and $0.2$, with $2N+1=101$ and $201$ replicas (indistinguishable); vertical lines mark $\omega=v_FQ$.
(d,e) Quasienergy spectra along $k_x-K_x$ at $Q=0.1$ for $\omega=0.40$ and $0.22$ ($\eta=+1$); red discs and blue intensity give the static weight, thin black lines the continuum hyperbolas with $m'_{\rm eff}$.
(f) Chern number of the lower band over the full Brillouin zone for $Q=4\pi/(\sqrt3\,90a)=0.081$ and both helicities $\eta=\pm 1$ ($300\times300$ grid);
only points where the band is identified over the whole zone are shown. Dashed line: $\omega_c$ from Eq.~(\ref{eq:2D-closing-numerical}).
}
\label{fig:honey}
\end{figure*}

For a single continuum Dirac cone, the sign change of $m_{\rm eff}'$ changes
its regularized Hall contribution by one quantum, but the absolute Chern
number is not fixed without an ultraviolet completion.  In a lattice material
the integer invariant must therefore be evaluated over the full Brillouin
zone using Eq.~(\ref{eq:TKNN}), including all valleys and bands.  Whether the
gap closing in Eq.~(\ref{eq:2D-gap-closing}) produces a net Chern transition or
a valley Hall transition depends on how the physical gauge wave couples to
the different Dirac cones; for the honeycomb lattice this is answered in Subsection~\ref{sec:lattice}: 
the two valleys see opposite helicities, so that a real gauge wave produces a net Chern transition.  
This distinction is especially important for
graphene, whereas a single surface Dirac cone of a three-dimensional
topological insulator has a different ultraviolet completion.

\subsection{Lattice model: honeycomb}
\label{sec:lattice}
To see which of these results survive on a lattice, we repeat the analysis for the honeycomb lattice 
with nearest-neighbor hopping $t$ and a staggered sublattice potential $\pm m$~\cite{Semenoff1984}. 
In the honeycomb lattice, the two Dirac cones of opposite chirality $\tau=\pm1$ are located at the corners of the Brillouin zone, $\bs{K}$ and $\bs{K}'$, with Fermi velocity $v_F=3ta/2$, bandwidth $6t$, and Chern number $C=0$ for $m\neq0$.
The circular gauge wave~(\ref{eq:A2D}) enters through the Peierls phases of the three bonds, expanded in Bessel functions so that all harmonics are kept
\cite{OkaAoki2009}.
Near each valley the first harmonic reduces to Eq.~(\ref{eq:V2D}) with $g=2v_FJ_1(A)\simeq v_FA$ and with the helicity seen by the valley equal to $\tau\eta$,
while the zeroth harmonic merely rescales $t$ by $J_0(A)$ and generates no static mass.
Equation~(\ref{eq:meff2D}) therefore becomes valley dependent, $m'_{\rm eff}(\tau)=m-\tau\eta\gamma_t^2g^2/\omega$:
a wave of helicity $\eta=+1$ inverts the mass at $\bs{K}$ only and $\eta=-1$ at $\bs{K}'$ only, and the band insulator is driven into a Floquet Chern insulator with $C=\eta$,
the propagating-wave analog of the Floquet--Haldane mass induced by circularly polarized light~\cite{OkaAoki2009,PhysRevB.84.235108}.

Figure~\ref{fig:honey} shows the results for 
the honeycomb lattice with parameters selected to be in the continuum regime:
We select $t=a=1$, $v_F=1.5$, $m=0.02$,
and the dimensionless Peierls amplitude $eAa=0.04$, which corresponds to $g=0.060$, $g/m=3$.
Note that this is of the same order as in Fig.~\ref{fig:dirac2d}.
In this regime of $\omega,g,m\ll W$ and $Q\ll1/a$, the lattice is expected to be irrelevant.
The gap at $\bs{K}$ for $\eta=+1$ [Fig.~\ref{fig:honey}~(a)] closely resembles that in Fig.~\ref{fig:dirac2d}~(a): the mass is inverted below a time-like closing line,
which follows Eq.~(\ref{eq:2D-closing-numerical}) to within about $1\%$ up to $\gamma_t^2=2.1$ at $Q=0.2$.
At $\bs{K}'$ [Fig.~\ref{fig:honey}~(b)] the same wave leaves the time-like regime gapped and inverts the mass only on the space-like side, along Eq.~(\ref{eq:spacelike-closing}).

The spectra along $k_x$ through $\bs{K}$ [Figs.~\ref{fig:honey}~(d,e)] show the dressed cone above the closing line and the inverted cone hybridized with its replicas below it, as in Fig.~\ref{fig:dirac2d-spectra}.
The lattice allows the Chern number to be evaluated over the full Brillouin zone [Fig.~\ref{fig:honey}~(f)]:
Eq.~(\ref{eq:TKNN}) with the lower band filled, i.e., the $s$-averaged curvature~(\ref{eq:BerryOmega}) integrated over the zone with the band identified by its static weight as in Sec.~\ref{sec:2D-numerics}.
It is quantized and changes from $C=0$ above the closing line to $C=+1$ ($-1$) below it for $\eta=+1$ ($-1$), at the $\omega_c$ of Eq.~(\ref{eq:2D-closing-numerical});
this identifies the Berry-flux jump of Fig.~\ref{fig:dirac2d}~(c) as a change of the Hall conductance by $e^2/h$ and fixes its absolute value.
Close to the light line, and where the one-photon resonance ring $2\ve(\bs{k})=\omega$ hybridizes the band strongly, the band cannot be followed over the whole zone and $C$ is not defined.

A feature that the lattice removes is the light-line singularity, which originates from the dense 
overlaps of the replica bands. 
This is seen in the replica participation number $N_{\rm rep}$ defined as
\begin{align}
N_{\rm rep}=\frac{1}{\sum_n w_n^2},\; w_n=\langle\Phi^n_{\alpha\bs K}|\Phi^n_{\alpha\bs K}\rangle,
\end{align}
which satisfies $\sum_n w_n=1$, evaluated for the state closest to $\ve=0$ at $\bs{K}$ [Fig.~\ref{fig:honey}~(c)].
It grows as the light line is approached and reaches $10$. 
In the continuum this quantity is expected to diverge, whereas on the lattice it is bounded and saturates as the cutoff is 
increased because the curvature of the band terminates the sequence of resonances described next.
In lattice systems, the condition for resonance between the original band and its $n$-th replica is given by ($\ve(\bs k)=\sqrt{m^2+t^2|f(\bs k)|^2},\; f(\bs k)=\sum_{j=1}^{3}e^{i\bs k\cdot\bs\delta_j}$)
\begin{align}
n\omega=\ve(\bs K+n\bs Q)\pm\ve(\bs K)
\label{eq:chord}
\end{align}
where the minus (plus) sign describes the crossing of the state $+\ve(\bs K)$ ($-\ve(\bs K)$) at $\bs K$ with the $n$-th replica of the upper band, i.e., an intraband (interband) resonance.
The multiple curves seen in Fig.~\ref{fig:honey}~(c) correspond to the condition Eq.~(\ref{eq:chord}) for different $n$s. 
In the continuum the resonances~(\ref{eq:chord}) accumulate at the light line for $n\to\infty$ and produce the singularity.


\section{Application to three-dimensional Dirac fermions and emergent Weyl points}
\label{sec:3D}

We now apply Eq.~(\ref{eq:HamDirac}) to a transverse circular wave propagating
in the $x$ direction,
\begin{align}
V_{\rm 3D}=\frac{A}{2}\left(\Gamma^y-i\Gamma^z\right).
\label{eq:CPL}
\end{align}
Here $A$ includes the charge, Fermi velocity, and vector-potential amplitude,
and therefore has dimensions of energy.  We denote the induced axial
energy--momentum four-vector by $b_{5{\rm eff}}^\mu$; its spatial part enters
the Hamiltonian as $\bs{\Gamma}\cdot\bs{b}_{5{\rm eff}}\,\gamma^5$ [cf.\ Eq.~(\ref{eq:HamDirac})].  This geometry can represent a
circularly polarized electromagnetic field~\cite{yoshikawa2022lightinducedchiralgaugefield}
or a coherent chiral phonon.  Because the $y$ and $z$ components are
transverse to the boost, their vector-potential amplitudes are unchanged in
the time-like rest frame; the enhancement at leading order is therefore the
frequency-redshift factor $\gamma_t$ in Eq.~(\ref{eq:1st}).

\begin{figure*}[tb]
\centering
\includegraphics[width=0.65\textwidth]{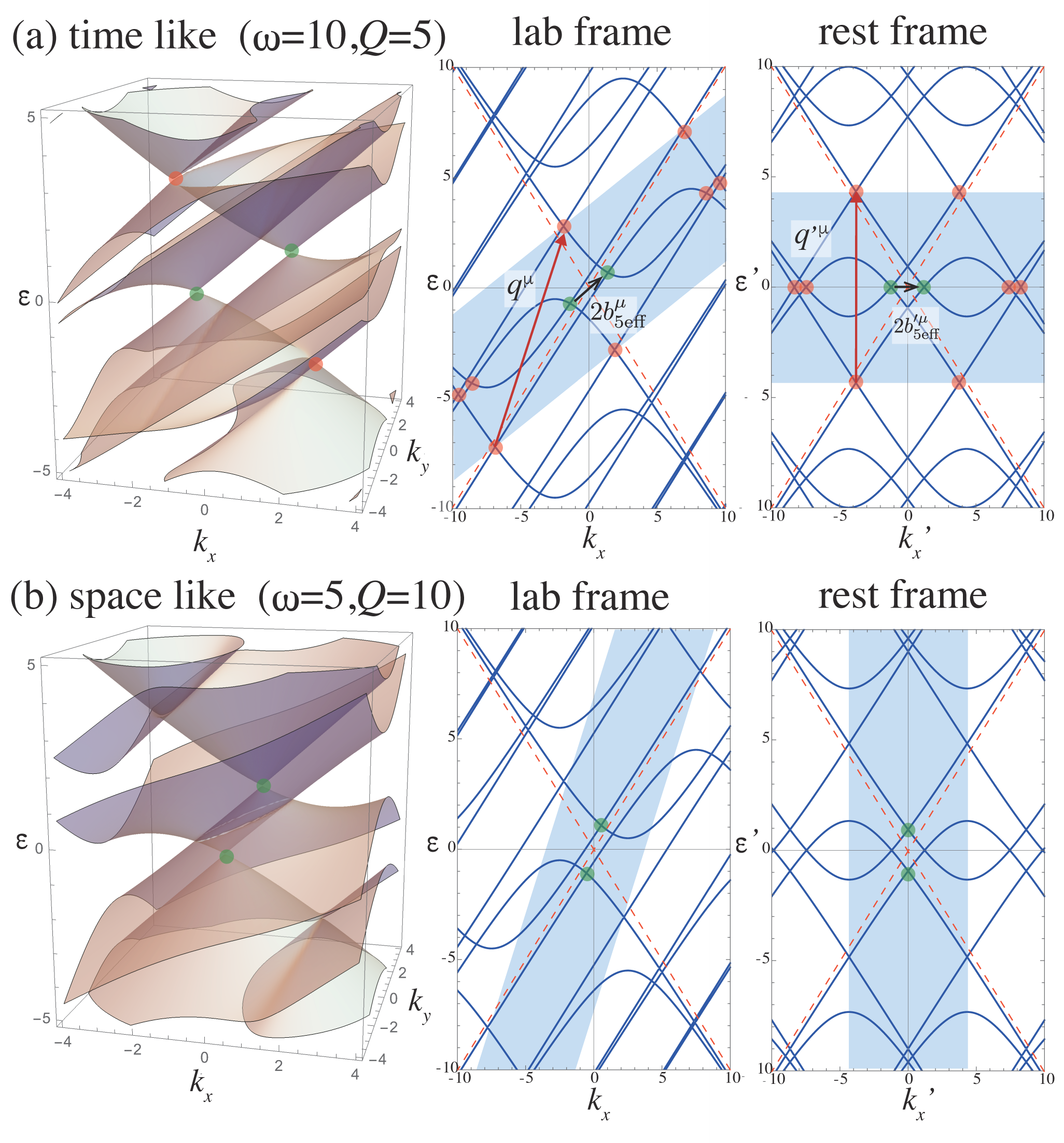}
\caption{
Floquet quasienergy spectra of massless Dirac electrons in the circularly polarized wave of Eq.~(\ref{eq:CPL}) with $A=3$ for the (a) time-like ($\omega=10,\; Q=5$)
and (b) space-like ($\omega=5,Q=10$) cases.
Left panels:
Spectra plotted as a function of  $k_x$  and  $k_y$  (with  $k_z = 0$).
Middle panels: Corresponding slices at  $k_y=k_z=0$. 
Right panels: Rest-frame spectra obtained via Lorentz transformation. 
The red dashed lines represent the original Dirac dispersion, while 
green dots denote the Weyl points created from the original Dirac node [in (b), the energy-split pair at $k'_x=0$, $\ve'=\pm0.94$ (Sec.~\ref{sec:3D-massive})], 
and red dots are Floquet-induced Weyl points originating from replica band hybridization. 
Blue shaded areas indicate the first Floquet--Bloch zones, $|q\cdot k|\le|q\cdot q|/2$, the boost of the rest-frame zones shown in the right panels.
}
\label{fig:Weylpoints}
\end{figure*}

\subsection{Massless Dirac fermion and emergent Weyl points}
\label{sec:3D-massless}
We consider the massless case ($m=0$) in the absence of external gauge fields $A_\mu,\;A_{5\mu}=0$.
The Floquet--Bloch spectra for time-like and space-like waves
(Fig.~\ref{fig:Weylpoints}) display the band tilting, stretching, and
Floquet--Bloch-zone distortion generated by the relation $k=a^{-1}k'$ between
the laboratory and rest frames.

A circularly polarized wave [Eq.~(\ref{eq:CPL})] breaks time-reversal symmetry and can generate Floquet Weyl points.
The green dots in Fig.~\ref{fig:Weylpoints}~(a) correspond to Weyl points pair-created from the original Dirac node~\cite{Wang_2014,PhysRevB.93.155107,PhysRevLett.116.026805,Hubener2017,yoshikawa2022lightinducedchiralgaugefield}, 
and the red dots represent emergent Weyl points originating from replica band hybridization~\cite{PhysRevB.96.041126,PhysRevResearch.6.L012027}.
In Fig.~\ref{fig:Weylpoints}~(b) the green dots mark the energy-split Weyl points due to the space-like wave. 
The momentum-space displacement of these Weyl points originates from the induced chiral gauge field~\cite{PhysRevB.93.155107} in the effective Floquet Hamiltonian~\cite{PhysRevB.84.235108}. 
To leading order in the high-frequency expansion [Eq.~(\ref{eq:1st})], 
this takes the form
\begin{align}
\bs{b}'_{5{\rm eff}}=\left(\gamma_t \frac{A^2}{\omega},0,0\right).
\label{eq:CGF1}
\end{align}
Applying the Lorentz transformation [Eq.~(\ref{eq:Lorentzfield})], the effective chiral gauge field in the lab frame is obtained as
\begin{align}
b_{5{\rm eff}}^\mu
=\frac{A^2}{\omega}
\left(\gamma_t^2\frac{v_F}{v},\gamma_t^2,0,0\right).
\label{eq:CGF2}
\end{align}
Near the shockwave limit ($v \to v_F$), the induced chiral gauge field is
strongly enhanced, and could be detected through optical
measurements~\cite{yoshikawa2022lightinducedchiralgaugefield}.
Furthermore, Eq.~(\ref{eq:CGF2}) predicts the generation of
$b_{5{\rm eff}}^0$, corresponding to a chiral chemical potential.
This component can, in principle, contribute to a chiral magnetic
response~\cite{CME}; for a chiral chemical potential $\mu_5$ the ideal
continuum expression is
\begin{align}
\bs{J}=\frac{e^2}{2\pi^2}\mu_5\bs{B},
\end{align}
although the measurable dc response depends on relaxation and on how the
nonequilibrium distribution is established; for a static $\mu_5$ in thermal equilibrium the current vanishes~\cite{VazifehFranz2013},
so that a nonzero response relies on the nonequilibrium occupation of the Floquet--Bloch states.

\begin{figure*}[tb]
\centering
\includegraphics[width=0.8\textwidth]{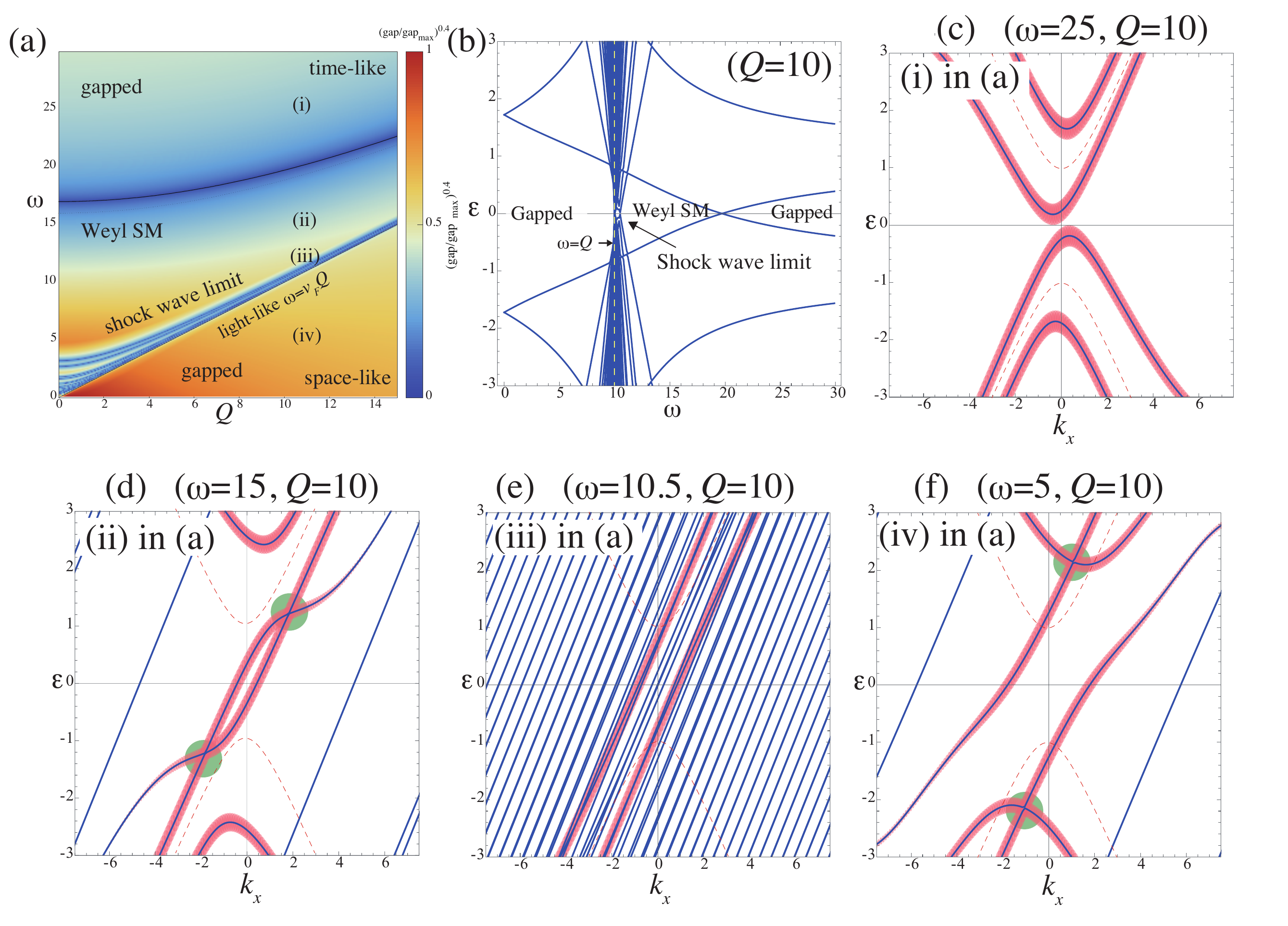}
\caption{
(a) Floquet phase diagram of a massive Dirac electron ($m = 1,\; A = 4$) driven by a circularly polarized wave [Eq.~(\ref{eq:CPL})]. 
The color represents the gap size at $\bs{k} = 0$, $(\Delta/\Delta_{\max})^{0.4}$, on the same scale as Fig.~\ref{fig:dirac2d}.
The gap closes on the hyperbola $\omega^2-v_F^2Q^2=(A^2/m+m)^2$ [Eq.~(\ref{eq:exactclosing}), solid line; the dotted line is the leading-order criterion $|b'_{5{\rm eff},x}|=|m|$],
marking a transition to a Floquet Weyl semimetal.
(b) Gap evolution with driving frequency $\omega$ at fixed $Q = 10$.
The system transitions from a gapped state [(i) in (a), panel (c)] to a Weyl semimetal [(ii), panel (d)] 
across $\omega_c=\sqrt{v_F^2Q^2+(A^2/m+m)^2}$.
Further reducing $\omega$ leads to a shockwave-like state with multiple band foldings near $\omega=v_FQ$ [(iii), panel (e)] and then to the space-like regime [(iv), panel (f)].
(c)--(f) Floquet-Bloch spectra at the points (i)--(iv) marked in (a), showing the gap closing, 
Weyl node emergence (green dots), and Floquet-induced spectral weight (red shading). 
The red dashed lines represent the original Dirac dispersion.
The red color in panels (c)--(f) represents the static weight 
$\langle\Phi^0_{\alpha\bs{k}}|\Phi^0_{\alpha\bs{k}}\rangle$.
}
\label{fig:phasediagram}
\end{figure*}

\subsection{Massive Dirac fermion and phase transitions}
\label{sec:3D-massive}
We now examine Floquet-induced phase transitions in massive Dirac electrons. 
First, in the undriven case, the Dirac mass $m$ opens a gap at the Dirac point, and the system is a trivial insulator. 
When the circularly polarized wave is applied, the Floquet-induced chiral gauge field $b_{5{\rm eff}}^\mu$ competes with the mass term, 
leading to a topological phase transition. 
This transition is characterized by the emergence of Weyl points in the Floquet--Bloch spectrum, 
which are separated in momentum space along the propagation direction of the wave.
The Floquet-induced chiral gauge field $b_{5{\rm eff}}^\mu$ becomes stronger as the driving frequency $\omega$ is lowered 
and approaches the shockwave limit. 
This is seen in Fig.~\ref{fig:phasediagram}
where the gapped state at large $\omega$ [(i), panel (c)] transitions to a Weyl semimetal [(ii), panel (d)] across the gap closing line.
Further lowering $\omega$ leads to a shockwave-like state [(iii), panel (e)] with multiple band foldings near $\omega=v_FQ$ 
and then to the space-like regime [(iv), panel (f)].

In the time-like case, the low-energy Floquet--Bloch spectrum for large $\omega'$ in the rest frame is
\begin{align}
\ve'_\alpha(\bs{k}')=\pm\Big\{&
\left[\sqrt{v_F^2(k_x')^2+m^2}
\pm b'_{5{\rm eff},x}\right]^2
\nonumber\\[-2pt]
&+v_F^2[(k_y')^2+(k_z')^2]\Big\}^{1/2}.
\label{eq:energymassive}
\end{align}
The Floquet-induced chiral field 
$b'_{5{\rm eff},x}=A^2/\omega'$ competes with the mass:
once it exceeds $|m|$, the Dirac gap closes and the node splits into a pair of Weyl nodes on the propagation axis,
signaling a transition into a Floquet Weyl semimetal~\cite{
Wang_2014,PhysRevB.93.155107,PhysRevLett.116.026805,Hubener2017,yoshikawa2022lightinducedchiralgaugefield,PhysRevB.96.041126,PhysRevResearch.6.L012027}.
On the axis, $k'_y=k'_z=0$, this picture can be made exact.
There the circularly polarized field in the rest frame simply rotates about the $x$ axis, and in a frame that co-rotates with the field the problem becomes static, so that the quasienergies follow for any $\omega'$ (Appendix~\ref{app:exact}):
\begin{align}
\ve'_\pm(k'_x)=\pm\sqrt{\Big(\sqrt{v_F^2k_x'^2+m^2}\mp\frac{\omega'}{2}\Big)^2+A^2}+\frac{\omega'}{2}\quad({\rm mod}\ \omega').
\label{eq:exactaxis}
\end{align}
The two signs are independent and give the four bands. 
For $\omega'\gg A,m$, this reduces to Eq.~(\ref{eq:energymassive}).
The gap at $\bs{k}'=0$ closes at
\begin{align}
\omega'_c=\frac{A^2+m^2}{m},\qquad {\rm i.e.}\qquad \omega^2-v_F^2Q^2=\Big(\frac{A^2}{m}+m\Big)^2,
\label{eq:exactclosing}
\end{align}
slightly above the leading-order estimate $|b'_{5{\rm eff},x}|=|m|$, i.e., $\omega'=A^2/m$.
In the $(Q,\omega)$ plane Eq.~(\ref{eq:exactclosing}) is a hyperbola, as required by the Lorentz invariance of $q^2$
[solid line in Fig.~\ref{fig:phasediagram}~(a); the dotted line is the leading-order estimate].
Below the gap closing line, the two Weyl nodes lie at $k_x'$ solving $\sqrt{v_F^2k_x'^2+m^2}=\omega'/2-\sqrt{\omega'^2/4-A^2}$, which reduces to $\sqrt{v_F^2k_x'^2+m^2}=b'_{5{\rm eff},x}$ for $\omega'\gg A$.
The other root, with $+\sqrt{\omega'^2/4-A^2}$, gives the replica-induced Weyl points [red dots in Fig.~\ref{fig:Weylpoints}~(a)].
Finally, while the high-frequency expansion predicts that $b'_{5{\rm eff},x}$ grows without bound as $\omega'\to0$, the exact quasienergies at $\bs{k}'=0$ tend to $\pm\sqrt{A^2+m^2}$:
the divergence in the shockwave limit is an artifact of the high-frequency expansion.

In wave-induced topological transitions, the bands are tilted. This gives rise to type-I and type-II Weyl points, 
which are distinguished by the slope of the energy dispersion at the Weyl point.  
In the laboratory frame, the Weyl cones are initially type II~\cite{Soluyanov2015} at the transition
and eventually tilt back to type I.  This can be understood from the
rest-frame energy slope
$d\ve'_\alpha/dk'=v_F\sqrt{1-(m/b'_{5{\rm eff},x})^2}$, which vanishes at
gap closing. 
The boost with velocity $u=v_F^2/v$ over-tilts the cone as long as this slope is smaller than $u$ [Fig.~\ref{fig:phasediagram}~(d)].
Going away from the transition, the slope increases and the cone tilts back to type I.

Close to the shockwave limit, the bands become densely folded and the
dynamics of a slowly varying gapped Hamiltonian becomes nearly
adiabatic~\cite{Berry1984}.
In the adiabatic limit, the rest-frame state is $\ket{\Psi'_{n\bs{k}}(t')}=\frac{1}{\sqrt{N}}e^{-i\ve'_n t'}\ket{n(\bs{k}+e\bs{A}(t'))}$
with the adiabatic basis satisfying $h(\bs{k})\ket{n(\bs{k})}=E_n(\bs{k})\ket{n(\bs{k})}$ with $h(\bs{k})=v_F\bs{\Gamma}\cdot \bs{k}+m\gamma^0$
and $n=1,\ldots 4$.
The quasi-energy is given as
$
\ve'_{\bs{k}}=\frac{1}{T'}\int_0^{T'}E_n(\bs{k}+e\bs{A}(t'))dt'-\gamma_B/T'$,
with the Berry phase $\gamma_B=i\oint dt'\,\bra{n}\pa_{t'}\ket{n}$ [which equals $-\gamma_{\alpha\bs{k}}$ of Eq.~(\ref{eq:Berryphase}), since $s=-\omega't'$ in the rest frame]. For $\bs{k}=0$, Eq.~(\ref{eq:exactaxis}) gives the exact quasienergies
$\ve'=\pm\sqrt{A^2+(m\mp\omega'/2)^2}+\omega'/2$, where the two signs are independent; for the two states near $+\sqrt{A^2+m^2}$, which are degenerate at $A=0$, the expansion for small $\omega'$ gives
$\ve'\simeq\sqrt{A^2+m^2}-\frac{\omega'}{2}\big(1\mp\frac{m}{\sqrt{A^2+m^2}}\big)$ (mod $\omega'$), which identifies
$\gamma_B=\pi\big(1\mp\cos\vartheta\big)$ (mod $2\pi$), $\cos\vartheta=m/\sqrt{A^2+m^2}$,
i.e., half of the solid angle $2\pi(1\mp\cos\vartheta)$ enclosed by the field $(m,A\cos\varphi',A\sin\varphi')$, 
which rotates clockwise about the $x$ axis, $\varphi'=-\omega't'$.  Thus,
the two Dirac states that are degenerate at $A=0$ acquire different Berry phases.
These expressions do not diverge in
the shockwave limit and connect smoothly to the Volkov solution at
$v=v_F$~\cite{Wolkow1935berEK}.  For stronger fields or faster driving,
interband tunneling~\cite{HiguchiLightCurrent} becomes important.  With a
circularly polarized wave, such tunneling is described by the twisted
Landau--Zener problem and can generate a nonuniform excitation distribution
and a current~\cite{Takayoshi2021}.  

In the space-like case, the gap at $\bs{k}=0$ remains open, while the tilted bands exhibit a chiral
flow [Fig.~\ref{fig:phasediagram}~(f)], analogous to the one-dimensional
case~\cite{yang2024quantizedacoustoelectricfloqueteffect}.
In the rest frame the wave is a static helical gauge field of wave number $Q'$, which is removed on the axis $k'_y=k'_z=0$ by the screw transformation $e^{-iQ'x'\Sigma_x/2}$ (Appendix~\ref{app:exact}).
In terms of the screw momentum $p=k'_x\mp Q'/2$, the exact spectrum on the axis is
$\ve'=\pm\sqrt{(\sqrt{v_F^2p^2+A^2}\mp v_FQ'/2)^2+m^2}$, which is the space-like counterpart of Eq.~(\ref{eq:exactaxis}) with $\omega'\to v_FQ'$.
Bands with different screw momentum do not hybridize on the axis, so that pairs of Weyl points split along the energy direction. 
The location of the energy-split Weyl points is
at $\ve'_W=\pm\sqrt{(\sqrt{v_F^2Q'^2/4+A^2}-v_FQ'/2)^2+m^2}$ with $\bs{k}'=0$. 
Their Lorentz boost gives the green dots in Fig.~\ref{fig:phasediagram}~(f).

\section{Possible realizations}
\label{sec:platforms}
The theory requires two ingredients: quasiparticles with a linear dispersion, which define the light line $\omega=v_FQ$,
and a propagating field whose phase velocity $v=\omega/Q$ can be brought close to $v_F$.
The first is met not only by Dirac and Weyl electrons~\cite{CastroNeto2009,Armitage2018} but also by Majorana fermions of Kitaev magnets~\cite{Kitaev2006,Banerjee2016},
spinons of Heisenberg chains~\cite{desCloizeaux1962,Lake2005}, and antiferromagnetic~\cite{Rezende2019} or Dirac magnons~\cite{Pershoguba2018,Chen2018CrI3}.
Their velocities range from $10^6$~m/s in graphene down to a few km/s in correlated systems,
where they can be tuned, e.g., by the twist angle of moir\'e materials~\cite{Cao2018} or are renormalized by Kondo screening~\cite{Lai2018,Dzsaber2021}.

For the second ingredient, Table~\ref{tab:drives} lists representative drives.
Light lies deep in the time-like regime with $\gamma_t\simeq1$.  
Thus, for example, the chiral chemical potential of Eq.~(\ref{eq:CGF2}) is suppressed relative to $b^x_{5{\rm eff}}$ by $v_F/v=nv_F/c\sim10^{-2}$.
Slow waves are therefore desired to induce the shockwave effects.
The evanescent fields of slow surface waves become circularly polarized in the quasistatic limit $v\ll c$ 
in the plane spanned by the propagation and decay directions with the helicity locked to the propagation direction~\cite{Bliokh2014}.
When the field decays normal to the surface, a three-dimensional Dirac or Weyl material below it experiences the rotating gauge field with a longitudinal component of Sec.~\ref{sec:2D}.
A two-dimensional layer needs a field that rotates within the layer; this is provided by waves that decay within the plane of the layer, by superposed strain waves, or by spin waves (Table~\ref{tab:drives}).
Velocity matching of this kind has already been discussed for Majorana fermions and phonons in Kitaev materials~\cite{Ye2022Majoranaphonon,Hauspurg2024}.

\begin{table*}[tb]
\caption{Representative propagating drives. $v$ is the phase velocity, to be compared with the quasiparticle velocity ($v_F\simeq10^6$~m/s in graphene, a few km/s in correlated systems). }
\label{tab:drives}
\begin{ruledtabular}
\begin{tabular}{llll}
Drive & Coupling & $\hbar\omega$ & $v$  \\
\hline
Circularly polarized light & electric field & $0.1$--$1.5$~eV & $c/n\gg v_F$  \\
Surface acoustic wave~\cite{Weiler2011,CerdaMendez2010} & piezoelectric field & $4$--$40~\mu$eV & $3$--$4$~km/s  \\
Rayleigh + shear-horizontal waves & strain (pseudogauge field) & $4$--$40~\mu$eV & $3$--$4$~km/s  \\
Surface plasmons, phonon polaritons~\cite{Ludeberg2017} & electric field & meV--$0.1$~eV & down to $\sim v_F$  \\
Edge plasmons, waveguide modes & electric field & meV--$0.1$~eV & down to $\sim v_F$ \\
Spin waves (TI surface)~\cite{Yokoyama2010} & exchange (vector potential) & $4~\mu$eV--meV & $0.1$--$10$~km/s  \\
\end{tabular}
\end{ruledtabular}
\end{table*}

The observables follow the two models studied here.
In two-dimensional Dirac systems such as graphene, moir\'e materials, or topological-insulator surfaces, the change of the Chern number (Secs.~\ref{sec:2D-numerics} and \ref{sec:lattice}) appears as a Hall or valley Hall response.
In three-dimensional Dirac and Weyl semimetals the induced chiral gauge field of Sec.~\ref{sec:3D} can be probed optically~\cite{yoshikawa2022lightinducedchiralgaugefield}.
In all cases the absorption of the wave itself is a direct probe, since for a linear dispersion the light line bounds the particle--hole continuum.

\section{Discussion}
\label{sec:discussion}

We have shown that propagating waves provide a qualitatively different route
to Floquet engineering from spatially uniform driving.  The space--time
reciprocal vector supplies both energy and momentum, and its Lorentz character
organizes the response into time-like, light-like, and space-like regimes.  A
time-like wave becomes a homogeneous, redshifted drive in its rest frame,
whereas a space-like wave becomes a static superlattice.  The approach to the
light line from the time-like side enhances virtual Floquet processes and,
once perturbation theory fails, produces a nonperturbative shockwave state
with strong replica mixing.

We have illustrated these ideas with two examples: a two-dimensional Dirac
electron in an in-plane circular gauge wave and a three-dimensional Dirac
electron in a transverse circular wave.  The former is relevant to graphene,
topological-insulator surfaces, and moir\'e materials, while the latter is
relevant to Dirac and Weyl semimetals.  In both cases the effective Lorentz symmetry 
of the Dirac equation allows the response to be understood in terms of the rest-frame drive, 
which is either a homogeneous drive or a static superlattice.  The shockwave 
enhancement of the Floquet engineering effects [Eq.~(\ref{eq:1st})] 
near the light line is the direct consequence. 

Several constraints delimit the shockwave regime.  The effective Lorentz
symmetry is only accurate within the linear Dirac window. In the lattice calculation
we showed that the light-line singularity is removed by the band curvature, 
which terminates the sequence of resonances.  
In addition, the nonequilibrium distribution of the Floquet--Bloch states is not addressed here; 
in a real system, heating, relaxation, and disorder become progressively important. 
Extending the analysis to systems with interactions and dissipation is an important direction for establishing
experimentally accessible steady states.

\section*{ACKNOWLEDGMENTS}
The author appreciates fruitful discussions with Christopher Yang, Iliya Esin,
Sota Kitamura, Swati Chaudhary, and Kush Saha. 
This work is supported by 
JSPS KAKENHI (No. JP23H04865, No. JP23K25837 and No. JP23K22487), MEXT, Japan, 
and JST CREST Grant No. JPMJCR19T3, Japan.

\bibliography{reference}

\appendix

\section{Floquet--Bloch theorem for a single propagating wave}
\label{app:FB}
The Hamiltonian $H(\bs{x},t)=H_0+V(\bs{x},t)$ with Eq.~(\ref{eq:pot1}) depends on $(\bs{x},t)$ only through the phase $q\cdot x=\omega t-\bs{Q}\cdot\bs{x}$
($H_0$ is translation invariant).
It is therefore invariant under all space--time translations $(\bs{x},t)\to(\bs{x}+\bs{d},t+\tau)$ with $q\cdot d=\omega\tau-\bs{Q}\cdot\bs{d}\in2\pi\mathbb{Z}$:
continuous translations that follow the wave, $(\tau,\bs{d})\propto(1,v\hat{\bs{Q}})$, which are Minkowski-orthogonal to $q^\mu$, continuous translations transverse to $\bs{Q}$, and discrete translations across the wave fronts.
These translations commute with each other and with the Floquet operator $i\pa_t-H$, so that the solutions can be chosen as their simultaneous eigenstates.
The eigenvalue of a translation with $q\cdot d=2\pi\ell$ is a phase $e^{-ik\cdot d}$, where $k^\mu=(\ve,\bs{k})$ is defined modulo $q^\mu$ [Eq.~(\ref{eq:FB-equivalence})];
writing $\ket{\Psi}=e^{-ik\cdot x}\ket{\Phi}$, the function $\ket{\Phi}$ is invariant under all these translations, i.e., it is a $2\pi$-periodic function of $q\cdot x$ alone,
which is Eq.~(\ref{eq:FBdecomposition}).
This is the statement of Bloch's theorem for the ``oblique space--time crystal'' generated by $q^\mu$~\cite{XuWu2018}.
Conversely, inserting Eq.~(\ref{eq:FBdecomposition}) into $i\pa_t\ket{\Psi}=H\ket{\Psi}$ and using
$H_0e^{i(\bs{k}+n\bs{Q})\cdot\bs{x}}=H_0(\bs{k}+n\bs{Q})e^{i(\bs{k}+n\bs{Q})\cdot\bs{x}}$,
$i\pa_t e^{-i(\ve+n\omega)t}=(\ve+n\omega)e^{-i(\ve+n\omega)t}$, and $V(\bs{x},t)=Ve^{-iq\cdot x}+V^\dagger e^{iq\cdot x}$ with $V=(V_1-iV_2)/2$,
the coefficient of $e^{-i(k+nq)\cdot x}$ gives
\begin{align}
(\ve+n\omega)\ket{\Phi^n}=H_0(\bs{k}+n\bs{Q})\ket{\Phi^n}+V\ket{\Phi^{n-1}}+V^\dagger\ket{\Phi^{n+1}},
\label{eq:FBapp}
\end{align}
which is Eq.~(\ref{eq:FBequation2}): $n$ counts the quanta absorbed from the wave, each carrying $(\omega,\bs{Q})$,
and $V$ ($V^\dagger$) absorbs (emits) one quantum.
The theorem holds for any band dispersion $H_0(\bs{k})$ and any number of internal degrees of freedom;
the relativistic dispersion enters only through the Lorentz transformation of Sec.~\ref{sec:Lorentz-framework}.
For $M$ waves with reciprocal vectors $q_1^\mu,\ldots,q_M^\mu$ the same argument gives $\ket{\Phi(q_1\cdot x,\ldots,q_M\cdot x)}$ 
and an $M$-dimensional Sambe index.

\section{Lorentz transformation}
\label{app:Lorentz}
The boost to the rest frame, $x'=ax$ (or $x'^{\mu}=a^\mu{}_\nu x^\nu$), is given by
\begin{eqnarray}
\left(
\begin{array}{c}
ct'\\
x'  \end{array}
\right)
=
\left(
\begin{array}{cc}
 \cosh\theta &   -\sinh\theta\\
-\sinh\theta &  \cosh\theta  
\end{array}
\right)
\left(
\begin{array}{c}
ct\\
x
  \end{array}
\right)
\label{eq:Lorentz}
\end{eqnarray}
with $y'=y,\; z'=z$. 

The chiral gauge field and electro-magnetic field in the two frames are related by the Lorentz transformation as ($c=v_F$)
\begin{equation}
    \left\{
    \begin{array}{l}
        c A'{}^5_0 = \gamma (c A^5_0 - \beta A^5_1) \\
       A'{}^5_x = \gamma (-\beta c A^5_0+ A^5_x) \\
        A'{}^5_y = A^5_y \\
        A'{}^5_z = A^5_z
    \end{array}
    \right.
\end{equation}
and 
\begin{eqnarray}
  &&  E'_x= E_x\\
&&    E'_y 
    = \gamma E_y - \gamma \beta c B_z\\
 &&   E'_z
    = \gamma E_z + \gamma \beta c B_y\\
 &&    cB'_x = cB_x\\   
 &&    cB'_y =  \gamma \beta E_z + \gamma c B_y\\
 &&    cB'_z     = -\gamma \beta E_y + \gamma c B_z.
\end{eqnarray}
Here $c=v_F$, $\beta=u/c=\tanh\theta$, and $\gamma=\cosh\theta$, 
with the boost velocity $u=v_F^2/v$ for a time-like wave and $u=v$ for a space-like wave 
[Eq.~(\ref{eq:Lorentz-boost})].

{\it Reciprocal vector and spectra.}---All scaled four-vectors of Eq.~(\ref{eq:scaled-four-vectors}) transform in the same way as $x^\mu$.
For the reciprocal vector, $q'=aq$ gives Eqs.~(\ref{eq:restframe1}) and (\ref{eq:restframe2}).
For the quasienergy and momentum, $k'=ak$; inverting this relation, a point $(\ve',k'_x)$ of the rest-frame spectrum appears in the laboratory at
\begin{align}
\ve=\gamma\left(\ve'+uk'_x\right),\qquad
k_x=\gamma\left(k'_x+\frac{u}{v_F^2}\ve'\right),
\label{eq:spectrum-boost}
\end{align}
with $\bs{k}_\perp$ unchanged.
This is how the rest-frame spectra in the right panels of Fig.~\ref{fig:Weylpoints} are mapped onto the laboratory spectra in the middle panels.

{\it Dirac spinor.}---Multiplying the Dirac equation that follows from Eq.~(\ref{eq:HamDirac}) by $\gamma^0/v_F$ puts it into the covariant form
\begin{align}
\left(i\gamma^\mu\pa_\mu-\frac{m}{v_F}\right)\Psi=0,\qquad \gamma^i\equiv\gamma^0\Gamma^i,
\end{align}
with $\pa_\mu=\pa/\pa x^\mu$ and $x^0=v_Ft$.
Under the boost the spinor transforms as $\Psi'(x')=D(a)\Psi(x)$, and the equation keeps its form provided
\begin{align}
D(a)^{-1}\gamma^\mu D(a)=a^\mu{}_\nu\gamma^\nu .
\label{eq:spinor-condition}
\end{align}
For a boost along $x$ this condition is satisfied by
\begin{align}
D(a)=e^{-\theta\Gamma^x/2}=\cosh\frac{\theta}{2}-\sinh\frac{\theta}{2}\,\Gamma^x .
\label{eq:Da}
\end{align}
To see this, note that $(\Gamma^x)^2=1$ and that $\gamma^0$ and $\gamma^1$ both anticommute with $\Gamma^x$, so that $\gamma^\mu D=D^{-1}\gamma^\mu$ for $\mu=0,1$.
Hence
\begin{align}
D^{-1}\gamma^0D&=\gamma^0D^2=\cosh\theta\,\gamma^0-\sinh\theta\,\gamma^1,
\nonumber\\
D^{-1}\gamma^1D&=\gamma^1D^2=\cosh\theta\,\gamma^1-\sinh\theta\,\gamma^0,
\end{align}
which are the two rows of Eq.~(\ref{eq:Lorentz}), while $\gamma^2$ and $\gamma^3$ commute with $\Gamma^x$ and are unchanged.
Since $\Gamma^x$ is Hermitian, $D(a)$ is Hermitian but not unitary, as is usual for the spinor representation of a boost.

{\it External fields.}---The coupling $\gamma^\mu A_\mu$ is covariant, $D^{-1}\gamma^\mu D\,A_\mu=\gamma^\nu A'_\nu$,
so that the rest-frame Hamiltonian has the form of Eq.~(\ref{eq:HamDirac}) with the transformed fields $A'^\mu=a^\mu{}_\nu A^\nu$.
A boost along $x$ leaves the transverse components $A^y$ and $A^z$ unchanged.
The transverse wave of Eq.~(\ref{eq:CPL}) therefore has the same amplitude in the rest frame.
This invariance concerns the field amplitudes, not the matrices: $\Gamma^y$ and $\Gamma^z$ anticommute with $\Gamma^x$ and hence do not commute with $D(a)$.
The in-plane wave of Eq.~(\ref{eq:A2D}), in contrast, has a longitudinal component, which mixes with the scalar potential as in Eq.~(\ref{eq:2D-boosted-potential}).

{\it Floquet--Bloch states.}---From Eq.~(\ref{eq:LorentzState}), the periodic parts of the Floquet--Bloch states in the two frames are related by
\begin{align}
\ket{\Phi(q\cdot x)}=D(a^{-1})\ket{\Phi'(q'\cdot x')},\qquad q'\cdot x'=q\cdot x .
\end{align}
The phase is a Lorentz scalar, so that the rest-frame state can equally be written as a function of $q\cdot x$, as in Eq.~(\ref{eq:LorentzState}).
Expanding both sides in $e^{-inq\cdot x}$ gives $\ket{\Phi^n}=D(a^{-1})\ket{\Phi'{}^n}$:
because $D(a^{-1})$ does not depend on $x$, it acts identically on every replica component, and the replica index $n$ is the same in the two frames.

\section{Exact solution on the propagation axis}
\label{app:exact}
On the propagation axis, $k'_y=k'_z=0$, the rest-frame problem of Sec.~\ref{sec:3D} can be solved exactly, because the circularly polarized drive is a rotation about the $x$ axis.
We use the matrix
\begin{align}
\Sigma_x={\rm diag}(\sigma_x,\sigma_x),
\end{align}
which generates rotations about the $x$ axis in spinor space: for any angle $\varphi$,
\begin{align}
e^{-i\varphi\Sigma_x/2}\,\Gamma^y\,e^{i\varphi\Sigma_x/2}=\Gamma^y\cos\varphi+\Gamma^z\sin\varphi .
\label{eq:rotation-identity}
\end{align}
The algebraic relations needed below are
\begin{align}
&\{\Gamma^x,\Gamma^y\}=\{\Gamma^x,\gamma^0\}=\{\Gamma^y,\gamma^0\}=\{\Gamma^y,\Sigma_x\}=0,
\nonumber\\
&[\Gamma^x,\Sigma_x]=[\gamma^0,\Sigma_x]=0,\qquad \Gamma^x\Sigma_x=\gamma^5 .
\label{eq:algebra}
\end{align}

{\it Time-like waves.}---In the rest frame the wave is homogeneous [cf.\ Eq.~(\ref{eq:2D-boosted-potential})],
\begin{align}
V'(t')=A\left[\Gamma^y\cos\varphi'+\Gamma^z\sin\varphi'\right],\qquad \varphi'=-\omega't'.
\end{align}
By Eq.~(\ref{eq:rotation-identity}) it is a rotated $\Gamma^y$, $V'(t')=A\,U(t')\,\Gamma^y\,U(t')^\dagger$, with
\begin{align}
U(t')=e^{-i\varphi'\Sigma_x/2}=e^{i\omega't'\Sigma_x/2}.
\end{align}
On the axis the Hamiltonian is $H'(t')=v_F\Gamma^xk'_x+V'(t')+m\gamma^0$.
The terms $\Gamma^yk'_y+\Gamma^zk'_z$, which would also be rotated, are absent there, and $\Gamma^x$ and $\gamma^0$ commute with $\Sigma_x$.
In the co-rotating frame, $\ket{\Psi'}=U\ket{\tilde\Psi}$, the Hamiltonian therefore becomes static,
\begin{align}
\tilde H=U^\dagger H'U-iU^\dagger\dot U=v_F\Gamma^xk'_x+A\Gamma^y+m\gamma^0+\frac{\omega'}{2}\Sigma_x .
\label{eq:Htilde-time}
\end{align}
To diagonalize it, we square it using Eq.~(\ref{eq:algebra}). All cross terms cancel except those of $\Sigma_x$ with $\Gamma^x$ and $\gamma^0$, so that
\begin{align}
\tilde H^2=v_F^2k_x'^2+A^2+m^2+\frac{\omega'^2}{4}+\omega'\,\big(v_Fk'_x\gamma^5+m\gamma^0\Sigma_x\big).
\end{align}
The two terms in the bracket anticommute, so the bracket squares to $v_F^2k_x'^2+m^2$ and has the eigenvalues $\pm\sqrt{v_F^2k_x'^2+m^2}$.
The eigenvalues of $\tilde H$ are thus
\begin{align}
\tilde E=\pm\sqrt{\Big(\sqrt{v_F^2k_x'^2+m^2}\mp\frac{\omega'}{2}\Big)^2+A^2},
\end{align}
where the two signs are independent.
Finally, the rotation is antiperiodic, $U(t'+T')=-U(t')$ with $T'=2\pi/\omega'$, which shifts the quasienergy by half a photon:
\begin{align}
\ve'=\tilde E+\frac{\omega'}{2}\quad({\rm mod}\ \omega').
\end{align}
This is Eq.~(\ref{eq:exactaxis}).
For $m=0$ the inner sign labels the chirality $\gamma^5=\pm1$, whose Weyl nodes shift in opposite directions;
for $m\ne0$ it labels the two states that are degenerate at $A=0$.

At $\bs{k}'=0$ the quasienergies are
\begin{align}
\ve'=\pm\sqrt{A^2+\Big(m\mp\frac{\omega'}{2}\Big)^2}+\frac{\omega'}{2},
\end{align}
for any $\omega'$. The gap around $\ve'=0$ closes when
\begin{align}
\sqrt{A^2+\Big(m-\frac{\omega'}{2}\Big)^2}=\frac{\omega'}{2},\qquad{\rm i.e.,}\qquad \omega'_c=\frac{A^2+m^2}{m},
\end{align}
which is Eq.~(\ref{eq:exactclosing}).
We have verified Eq.~(\ref{eq:exactaxis}) against the direct diagonalization of Eq.~(\ref{eq:FBequation2}) with up to $2N+1=521$ replicas.
The same two-level calculation for the two-dimensional model at $Q=0$ gives $\ve=\pm\sqrt{g^2+(m-\eta\omega/2)^2}+\omega/2$ and $\omega_c=(g^2+m^2)/m$, quoted in Sec.~\ref{sec:2D-numerics}.

{\it Space-like waves.}---In the rest frame the wave is a static helical field,
\begin{align}
V'(x')=A\left[\Gamma^y\cos Q'x'+\Gamma^z\sin Q'x'\right].
\end{align}
It is removed on the axis by the screw transformation
\begin{align}
U(x')=e^{-iQ'x'\Sigma_x/2},
\end{align}
which plays the role of $U(t')$ above.
The price is a shift of the kinetic term, $-i\pa_{x'}\to p-\frac{Q'}{2}\Sigma_x$, where $p$ is the screw momentum, so that by Eq.~(\ref{eq:algebra})
\begin{align}
\tilde H=v_F\Gamma^xp+A\Gamma^y+m\gamma^0-\frac{v_FQ'}{2}\gamma^5 .
\end{align}
The same algebra as for Eq.~(\ref{eq:Htilde-time}), with the roles of $A$ and $m$ interchanged, gives
\begin{align}
\tilde E=\pm\sqrt{\Big(\sqrt{v_F^2p^2+A^2}\mp\frac{v_FQ'}{2}\Big)^2+m^2}.
\end{align}
The screw momentum is related to the Bloch momentum by $k'_x=p\pm Q'/2$ (mod $Q'$) for the two eigenvalues of $\Sigma_x$.
Bands with different $p$ do not hybridize on the axis, so that their crossings are exact degeneracies.
In particular, at $k'_x=0$ ($p=\mp Q'/2$) the two sectors cross at
\begin{align}
\ve'_W=\pm\sqrt{\Big(\sqrt{\frac{v_F^2Q'^2}{4}+A^2}-\frac{v_FQ'}{2}\Big)^2+m^2}
\end{align}
for any $m$. These are the energy-split Weyl points of Fig.~\ref{fig:phasediagram}~(f) [and of Fig.~\ref{fig:Weylpoints}~(b) for $m=0$].
Away from the axis the screw symmetry involves a rotation of $(k'_y,k'_z)$, so that these crossings are isolated points in three dimensions;
numerically the splitting grows linearly with $k'_y$ (e.g., $0.146$ at $k'_y=0.1$ for $m=1$, $A=4$, and $Q'=5\sqrt3$).

{\it Static mass renormalization of the two-dimensional model.}---In the space-like rest frame of Sec.~\ref{sec:2D}, we write the Fourier matrix as
\begin{align}
V'=c+a\sigma_x+ib\sigma_y,\qquad a=\frac{\gamma_sg}{2},\quad b=\frac{\eta g}{2},\quad c=\frac{\gamma_s\beta_sg}{2},
\end{align}
where $c$ is the scalar potential generated by the boost.
The state at $\bs{k}'=0$ couples to the Bloch copies at $k'_x=+Q'$ through $\mathcal{H}^{1,0}=V'$ and at $k'_x=-Q'$ through $\mathcal{H}^{-1,0}=V'^\dagger$.
Their Hamiltonians are $\pm v_FQ'\sigma_x+m\sigma_z$, and second-order perturbation theory gives
\begin{align}
H_{\rm eff}=H^{(0)}_{\rm 2D}-\frac{V'^\dagger(v_FQ'\sigma_x+m\sigma_z)V'+V'(-v_FQ'\sigma_x+m\sigma_z)V'^\dagger}{v_F^2Q'^2+m^2}.
\end{align}
The products follow from
\begin{align}
V'\sigma_xV'^\dagger&=(a^2-b^2+c^2)\sigma_x+2ac+2bc\,\sigma_z,
\nonumber\\
V'\sigma_zV'^\dagger&=(c^2-a^2-b^2)\sigma_z-2bc\,\sigma_x+2ab,
\end{align}
and from the same expressions with $b\to-b$ for $V'^\dagger\cdots V'$.
The $\sigma_x$ terms cancel, and the coefficient of $\sigma_z$ is
\begin{align}
m'_{\rm eff}=m-\frac{-4v_FQ'bc+2m(c^2-a^2-b^2)}{v_F^2Q'^2+m^2}.
\end{align}
With $c^2-a^2-b^2=-g^2/2$ and $4v_FQ'bc=\eta\gamma_s\beta_sg^2v_FQ'$, this is Eq.~(\ref{eq:meff-spacelike}).
Without the scalar potential ($c=0$) the correction would be $(1+\gamma_s^2)g^2m/[2(v_F^2Q'^2+m^2)]$, independent of the helicity:
the mass inversion on the Bloch side is entirely due to the boost-generated scalar potential.

\end{document}